\documentclass[aps,prb,twocolumn]{revtex4-1}
\usepackage{amsmath}
\usepackage{amssymb}
\usepackage{epsfig}
\usepackage{graphics}

\begin{document}
\title{Quantum hexatic order in two-dimensional dipolar and charged fluids}
\author{Georg M. Bruun}
\affiliation{Department of Physics and Astronomy, University of Aarhus, Ny Munkegade, DK-8000 Aarhus C, Denmark}
\author{David R. Nelson}
\affiliation{Department of Physics, Harvard University, Cambridge, Massachusetts 02138, USA}
\begin{abstract}
Recent advances in cold atom experimentation suggest that studies of quantum two-dimensional melting of dipolar molecules, with dipoles
 aligned perpendicular to ordering plane, may be on the horizon.    An intriguing aspect of this problem is that two-dimensional \emph{classical} aligned dipoles (already studied in great detail in soft matter experiments on magnetic colloids) are known to melt via a two-stage process, with an intermediate hexatic phase separating the usual crystal and isotropic 
 fluid  phases.  We estimate here the effect of quantum fluctuations on this hexatic phase, for both dipolar systems and charged Wigner crystals.   Our approximate phase diagrams rely on a pair of Lindemann criteria, suitably adapted to deal with effects of thermal fluctuations in two dimensions.   As part of our analysis, we determine 
 the phonon spectra of quantum particles on a triangular lattice interacting with repulsive $1/r^3$ and 
$1/r$ potentials.  A large softening of the transverse and longitudinal phonon frequencies, due to both lattice effects and quantum fluctuations, plays a significant role in our analysis.   The hexatic phase is predicted to survive down to very low temperatures.\end{abstract}

\maketitle

The melting of crystals is a fundamental topic in condensed matter physics that has been studied for more than a century. Nevertheless,  
one  lacks a quantitative understanding of the melting for many materials. This deficiency is partly due to 
imprecise knowledge of the particle interactions that control  the relevant phase transitions on a microscopic scale. 
In two dimensions (2d), however, a defect-mediated theory of melting is available.   Building on pioneering work proposing an 
entropically-driven proliferation of dislocations by Kosterlitz and Thouless and by Berezinski,~\cite{Jose} 
and on unpublished work of Feynman,~\cite{Feynman} a defect mediated theory was worked out.~\cite{NelsonBook} The detailed theory actually invokes a 
\emph{sequential} unbinding of dislocations and disclinations, with the usual latent heat of a first order melting transition spread out 
over an intermediate hexatic phase.  The hexatic phase is characterised by extended bond  orientational 
order at intermediate densities or  temperatures. It  has been observed in 
a series of impressive experiments using colloidal particles with a diameter of 4.5$\mu$m and an induced magnetic moment, 
with dynamics confined to an air-water interface.~\cite{Gasser,Strandburg} Although this system, characterised by long range $1/r^3$ dipole-dipole 
interactions, is thoroughly understood  in the classical regime,  it is presently unknown whether the hexatic phase exists when 
quantum fluctuations play a major role.

The advancing field of ultra-cold   gases consisting of heteronuclear molecules with an electric dipole moment 
 promises to change this situation. The dipolar gases are expected to exhibit qualitatively 
 new physics, with several experimental groups  reporting impressive 
 progress towards achieving quantum degeneracy in these systems.~\cite{Wu,Ni,Chotia,Heo,Pasquiou} 
 Using dipolar gases, one should finally be able to probe experimentally the role of quantum fluctuations on the hexatic phase and study its stability
 at very low temperatures.  Recent Monte-Carlo calculations predict that a 2d dipolar gas with the moments perpendicular to the 2d plane 
 exhibits a quantum phase transition directly to a hexagonal crystal phase  at zero temperature 
 $T=0$.~\cite{Astrakharchik,Buchler} However, the effects of quantum zero point motion on the hexatic phase, accessible now  for the first time via cold dipolar gases, have
  not been discussed. 
  
In this paper, we explore this question by analysing the stability of  crystal and hexatic order of a  2d system of dipoles including
both thermal and quantum effects. First, we calculate the classical elastic coefficients of the crystal  from the phonon spectrum. 
We then show how quantum effects soften the crystal by decreasing these coefficients. Using  Lindemann criteria 
suitably modified to treat both 2d thermal fluctuations and quantum effects, 
we study the successive loss of translational and orientational order that lead to the melting of the crystal and hexatic phases. 
    The relevant Lindemann numbers are extracted in the classical regime as well 
  as for $T=0$ by comparing with Monte-Carlo and experimental results. 
   Throughout the paper, we construct a useful comparison between  the 2d dipolar system and a 2d fluid
  consisting of negatively charged particles immersed in a uniform background of positive charges -- the Jellium model.
An experimental realisation of charged systems in the classical limit with a Wigner crystal phase at low temperatures 
has been known for quite some time, in the form of a 2d electron 
gas trapped by a positively charged capacitor plate to the surface of liquid helium.~\cite{Grimes}   For computer simulation evidence that 
this system melts via a dislocation mechanism and may possess an intermediate hexatic phase, see Ref.~\onlinecite{Morf} and 
\onlinecite{Muto} respectively.
     We obtain essentially the same Lindemann numbers for the dipolar and the charged systems, which 
   suggests that the melting of these phases is a geometric phenomenon, insensitive  
  to the detailed form of the interaction potential. Similar conclusions resulted from Monte-Carlo simulations 
  of quantum hard sphere systems,~\cite{Runge} and from a meta-analysis of  Monte-Carlo results for the freezing of two-dimensional systems.~\cite{Babadi}
We show that  quantum effects initially \emph{increase} the temperature range where the 
  hexatic phase is stable when the  coupling strength is decreased from the strong 
  coupling classical regime, provided the temperature dependence of the Lindemann numbers can be neglected. 
   A tentative phase diagram is then provided showing that quantum effects are important
   even for very large interaction strengths where 
  one would naively expect the system to be deep in the classical regime. Finally, we discuss the possible experimental 
  observation of  the hexatic phase in the quantum regime using cold dipolar gases.

  We conclude this introduction with a few observations about the Lindemann criterion for melting of quantum and classical systems in 2d
   and the nature of quantum hexatics.  As originally proposed by Lindemann,~\cite{Lindemann} 
   one first calculates the mean-square displacement $W=\langle|{\mathbf u}({\mathbf r})|^2\rangle$
of a single particle away from its equilibrium lattice position, where $\langle\ldots\rangle$  represents an ensemble average. 
Melting as a function of, say, the temperature or density  then occurs when the root mean square displacement exceeds a fixed fraction of the lattice spacing $a$ , 
 i.e.\ for $\sqrt W\ge c_La$, where the Lindemann number is typically in the range $c_L\approx0.1-0.3$.~\cite{Blatter} 
 This rough criterion fails, however, in 2d classical solids, because $W$ diverges logarithmically with system size.
  Here we use an alternative formulation that focuses on the stretching of a nearest neighbor distance,~\cite{FisherFisher} namely 
 \begin{equation}
\Delta({\mathbf r})=\sqrt{\langle |{\mathbf u}({\mathbf r}+{\mathbf b})-{\mathbf u}({\mathbf r})|^2\rangle}\ge\gamma_m a
\label{Lindemann}
\end{equation}
where ${\mathbf b}$ connects nearest neighbor lattice sites ($|{\mathbf b}|=a$) and $\gamma_m$ is an alternative Lindemann number describing this new measure
 of the loss of translational order.    The quantity  $\Delta({\mathbf r})$ remains finite in the thermodynamic limit even in 2d and, as we show here, can also 
 be computed in the quantum regime all the way down to  $T=0$ for simple pair potentials.   Moreover, a related criterion allows us to estimate where the order associated with the
  \emph{rotational} broken symmetry of a 2d crystal is lost due to thermal or quantum fluctuations, 
  namely    
  \begin{equation}
\Delta_\theta({\mathbf r})=\sqrt{\langle\theta^2({\mathbf r})\rangle}=\frac 1 2 \sqrt{\left\langle \left|\partial_xu_y({\mathbf r})-\partial_yu_x({\mathbf r})\right|^2\right\rangle}\ge\gamma_i 
\label{LindemannHexatic}
\end{equation}
 where $\gamma_i$ is a Lindemann number for the loss of bond orientational order.   Here, $\theta({\mathbf r})=[\partial_xu_y({\mathbf r})-\partial_yu_x({\mathbf r})]/2$ 
 is the local phonon-induced twist of the crystallographic axes,~\cite{LandauLifshitz} a quantity whose  
 fluctuations are known to remain finite even in the limit of infinite system size for a 
 classical 2d crystal.~\cite{Mermin} Note that $\Delta({\mathbf r})\approx \sqrt{\langle[({\mathbf b}\cdot\nabla){\mathbf u}({\mathbf r})]^2\rangle}$
  has a similar gradient structure to  $\Delta_\theta({\mathbf r})$.
    These two different Lindemann numbers $\gamma_m$ and $\gamma_i$ 
    allow for two distinct melting temperatures, characterized by the successive loss of first translational and then orientational order,~\cite{NelsonHalperin}
     a scenario we \emph{know} occurs for classical colloidal particles interacting with repulsive long range $1/r^3$ dipole-dipole interaction.~\cite{Gasser}
Our evaluation of the criterion (\ref{LindemannHexatic}) using \emph{crystalline} phonon spectra to estimate the extent of the hexatic phase 
seems reasonable, provided local orientational order remains robust after long range translational order is lost, as is the case for the dislocation-disclination 
theory of classical 2d melting,~\cite{NelsonHalperin} and in situations where a weakly first order transition leads to a hexatic phase.~\cite{Engel}
 We also note that our use of phonon displacements from an underlying reference crystal implicitly treats quantum particles as distinguishable 
 (a similar approximation is used in the Debye theory of the specific heat of crystals~\cite{AshcroftMermin}), so we are effectively looking at the melting of quantum 
 particles with Boltzmannian statistics.~\cite{Clark}
    We assume that the exchange interactions that distinguish bosons from fermions play only a minor role in determining the locations of quantum melting transitions.
    
\begin{figure}
\epsfig{file=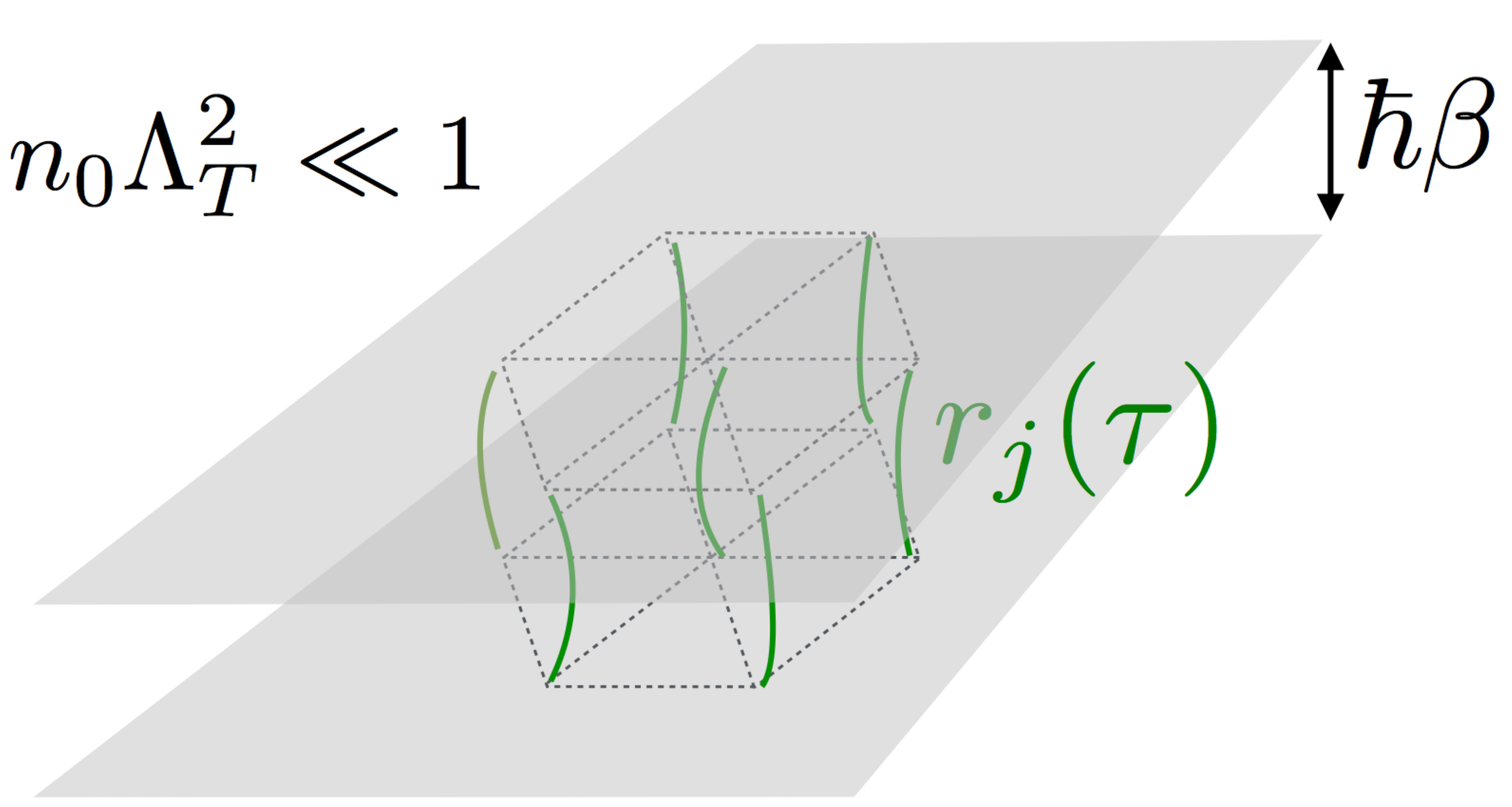,width=0.49\columnwidth}
\epsfig{file=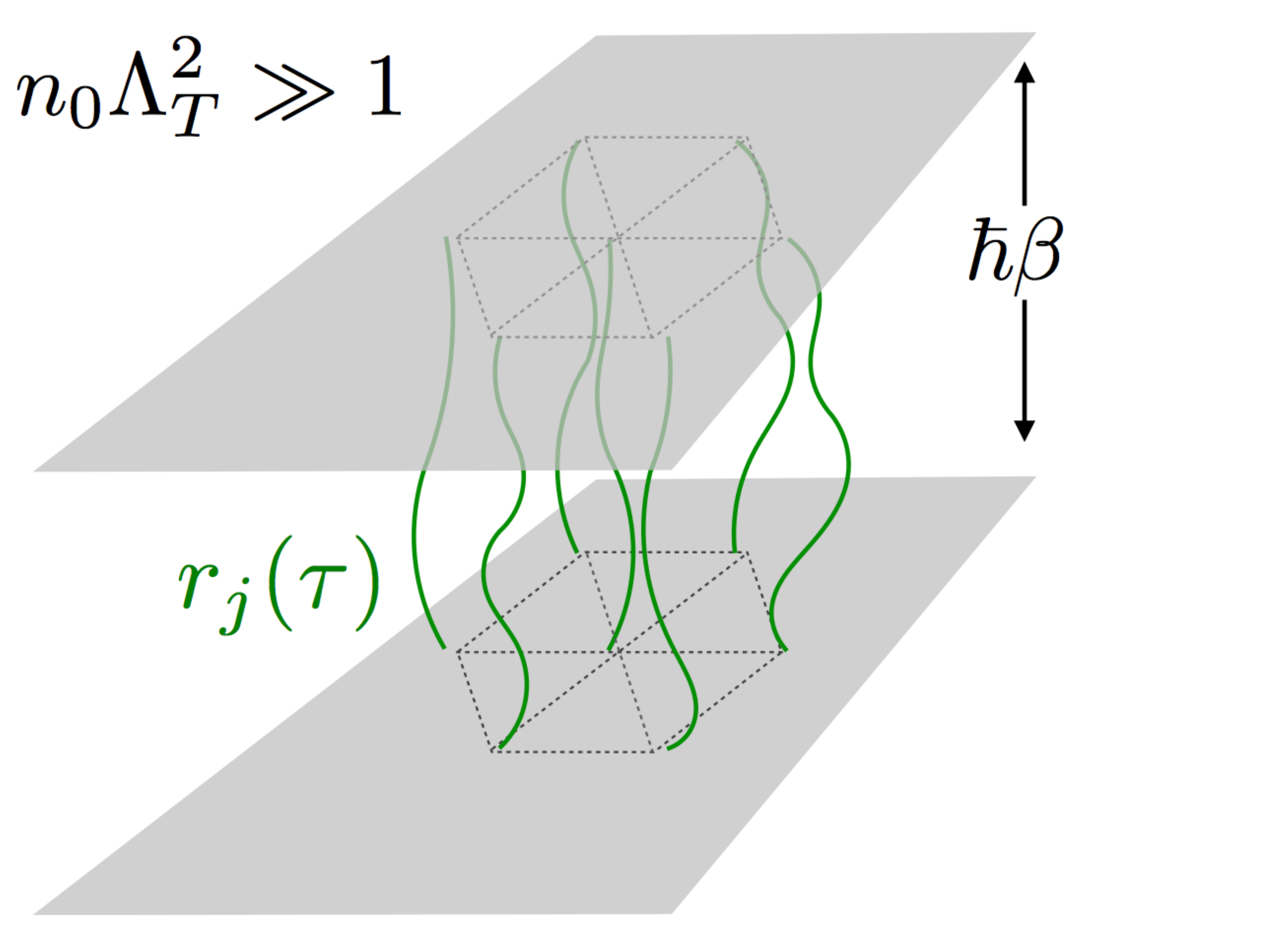,width=0.49\columnwidth}
\caption{The quantum partition function (\ref{Zpath}) is obtained by integrating over particle world  lines  in imaginary time $\tau$. 
Left: The  high temperature regime limit
$n_0\Lambda_T^2\ll 1$ where one recovers the classical 2d system. Right: The low temperature quantum regime where the problem becomes 2+1 dimensional.} 
\label{WorldlineFig}
\end{figure}

 These Lindemann inequalities are \emph{criteria}, and of course do not themselves constitute a \emph{theory} of quantum or classical melting.   
 We are not aware of a reliable microscopic theory of 2d quantum melting, and even the defect-mediated melting of classical particles in two dimensions could be preempted by a direct 
 first order transition from a crystal to an isotropic liquid.~\cite{NelsonBook}    
 It is also worth noting the rather different nature of classical as opposed to quantum melting in two dimensions. 
  This difference is particularly evident in the Feynman path integral formulation of nonrelativistic quantum statistical mechanics,~\cite{FeynmanHibbs}
   where classical particles are replaced by configurations of particle world lines in imaginary time, see Fig. \ref{WorldlineFig}.   
   We allow, for simplicity, only the identity permutation with periodic boundary conditions across an imaginary time slab of thickness  $\beta\hbar$.  
In the absence of interactions, these trajectories when projected down the imaginary time axis
behave like two dimensional random walks as a function of the imaginary time variable, with a size given by the thermal de Broglie 
wavelength $\Lambda_T=\sqrt{2\pi\hbar^2\beta/m}$. 
In the classical limit $\Lambda_T\ll n_0^{-1/2}$, where $n_0$  is the areal particle density, the particle world lines are short and nearly straight;  hence, the usual Lindemann picture 
for melting of point-like particles applies when  interactions are turned on.   However, in the highly quantum limit  $\Lambda_T\gg n_0^{-1/2}$, 
quantum and thermal fluctuations act on a crystal of long wiggling \emph{lines}:  as illustrated in Fig.\ \ref{WorldlineFig}, a particle world line ${\mathbf r}_j(\tau)$
that makes a large excursion within its confining cage at imaginary time $\tau$ 
is connected by the kinetic energy interaction $m\int_0^{\hbar\beta}|d{\mathbf r}_j(\tau)/d\tau|^2/2$  to time slices above and below, and hence can more easily recover and return
to its equilibrium position when the slab thickness is large.  It is harder to melt arrays of lines in $2 + 1$ 
dimensions than point-like particles in two dimensions with the same pair potential.    
 Thus, we should not be surprised if the Lindemann numbers $\gamma_m$ and $\gamma_i$ depend somewhat on $n_0\Lambda_T^2$, 
with larger Lindemann numbers required to produce melting when  $n_0\Lambda_T^2\gg 1$.   This is indeed what we find fitting to experiments on
colloids and quantum simulations of power law potentials, with, e.g., $\gamma_m$ ranging from $\sim0.1$  in the classical limit to $\sim0.3$ when quantum effects predominate.   
Phonon nonlinearities can give rise to a weak temperature-dependence of the long wavelength elastic constants, even in the absence of quantum fluctuations.~\cite{Morf}
  We neglect such effects here for simplicity.
  With these understandings, we believe the criteria sketched above can provide a rough map of where to look for quantum melting in the new arena provided by cold quantum gases.
  %%%%%%%%%%%%%%%%%%%%%%%%%%%%%%%%%%%%%%%%%%%%%%%%%%%%
%%%%%%%%%%%%%%%%%%%%%%%%%%%%%%%%%%%%%%%%%%%%%%%%%%%%

What would a quantum hexatic look like, if next generation experiments were to discover such a thing?  Roughly speaking, it would be a quantum liquid crystal, a cousin of the 
long-sought supersolid phase of $^4$He if the particles were bosons.~\cite{Reppy,Balibar}   A fermionic analog would be the quantum nematic studied by Oganesyan \textit{et.\ al.}
 in electronic systems,~\cite{Oganesyan} which exhibits a two-fold rather than six-fold anisotropy. These authors also considered the possibility of an electronic quantum 
 hexatic.  A hexatic quantum fluid would  display a fuzzy, six-fold-symmetric diffraction pattern, indicating extended orientational correlations, somewhat like a poorly averaged powder diffraction pattern.   However, \emph{unlike} a classical polycrystal, a quantum hexatic would be a fluid with zero shear modulus!  If composed of bosons, it could develop a nonzero superfluid density and support supercurrents at sufficiently low 
 temperatures.~\cite{Mullen} 
 The nature of hexatic order in real space, with extended correlations in the orientations of distant six-fold particle clusters, is discussed in Ref.\ \onlinecite{NelsonBook}.

\begin{figure}
\epsfig{file=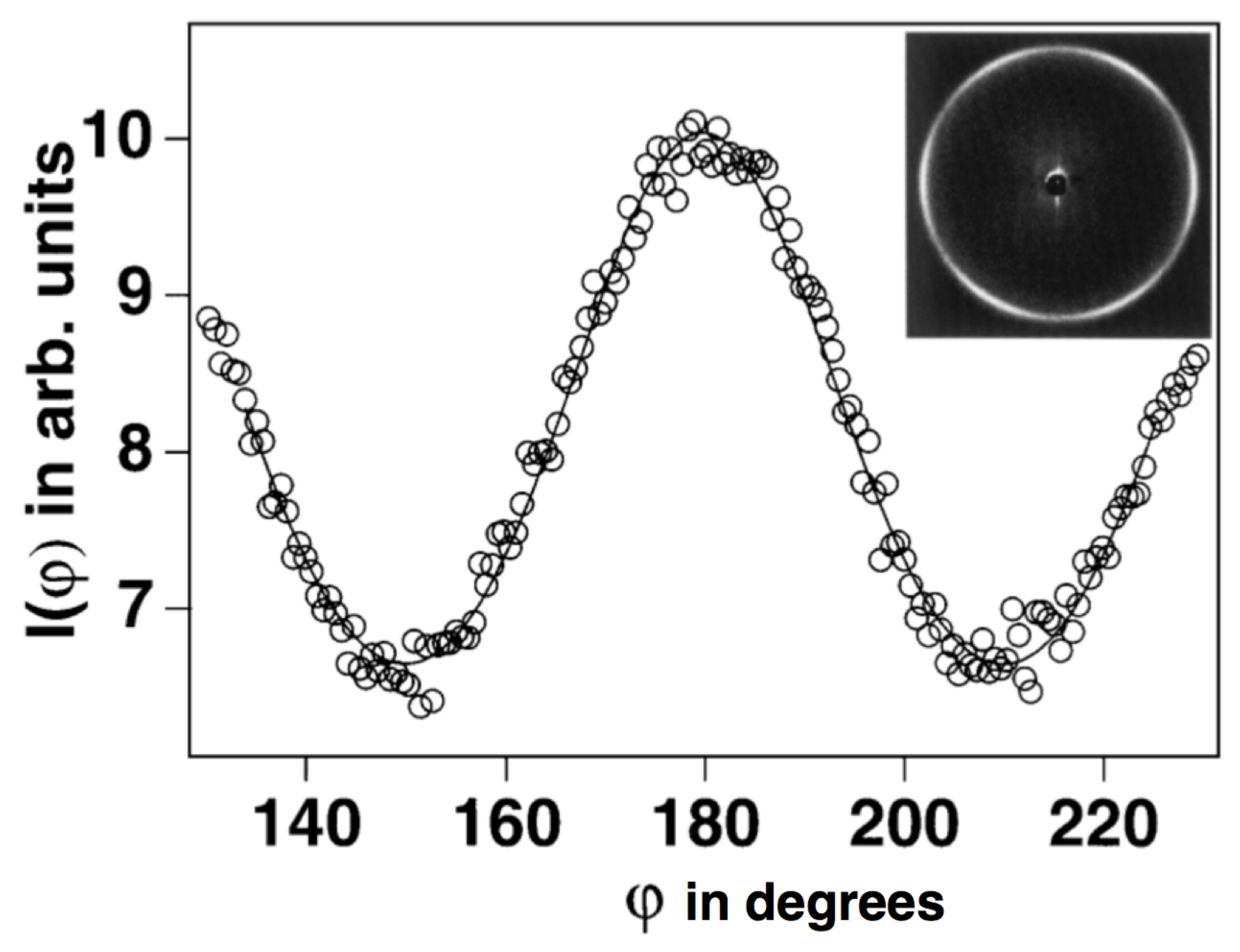,width=0.9\columnwidth}
\caption{X-ray diffraction pattern as a function of scattering angle $\phi$
from a sample of aligned DNA-molecules exhibiting the six-fold symmetry characteristic of the hexatic phase. Reprinted with 
permission from Ref.\ \onlinecite{Strey}.} 
\label{ExperimentalFig}
\end{figure}

In the crude phase diagrams constructed here, we have taken a conservative approach, and assumed the intermediate hexatic phase is squeezed out as 
$T\rightarrow0$, leaving behind a transition from a quantum solid directly to an isotropic quantum liquid.   But this need not be the case:   consider particles
 interacting with a screened, 2d Yukawa potential, 
 \begin{equation}
 V(r)=\epsilon_0K_0(\kappa r)
 \label{Vpair}
 \end{equation}
 where $K_0(x)$ is the modified Bessel function of the second kind, $K_0(x)\sim -\log x$ for $x\ll 1$, $K_0(x)\sim \exp(-\kappa x)$ for $x\gg 1$, and
 $\kappa^{-1}$ is a screening length.  Such a potential describes interactions between vortex lines with weak thermal fluctuations in Type II superconductors
 with an external magnetic field, where $\kappa^{-1}$
   is the London penetration depth.    When the lines are very long compared to the vortex line spacing, and pinning is negligible, the classical statistical mechanics of these
    three dimensional lines at finite temperatures can be mapped via the transfer matrix method onto the quantum statistical mechanics of 2d bosons at $T=0$
      interacting with the pair potential Eq.\ (\ref{Vpair}).\cite{NelsonSeung}     Here, the temperature $T$ of the 3d superconductor plays the role of $\hbar$ 
      and the thickness of the bulk superconductor plays the role of $\hbar\beta$
      in the equivalent 2d quantum system.  A dislocation loop unbinding model then leads directly to an entangled liquid of vortex lines, with long range six-fold bond orientational order, equivalent via the path integral mapping to a zero temperature quantum hexatic.\cite{Marchetti}   Although it has not yet been possible to check for hexatic order in melted vortex liquids in Type II superconductors (the signal from neutron diffraction is quite weak), something very like an entangled line hexatic has been seen in X-ray diffraction experiments off partially ordered arrays of aligned DNA molecules.~\cite{Strey} 
        When these charged, linear polymers are aligned by an external field, a screened, Debye-H\"uckel interaction arise in the perpendicular direction which has precisely 
the form (\ref{Vpair}).   The characteristic line hexatic diffraction pattern from this work is reproduced in Fig.\ \ref{ExperimentalFig}. 
  The 2d structure function $\langle|n({\mathbf q})|^2\rangle$ 
for  particles in a quantum hexatic at low temperatures should look very similar.    
    %%%%%%%%%%%%%%%%%%%%%%%%%%%%%%%%%%%%%%%%%%%%%%%%%%%%

\section{Lam\'e coefficients}
In this section, we calculate the elastic coefficients from the phonon modes of a 2d hexagonal crystal consisting of charged particles 
with a neutralising background charge density, or particles 
with a dipole moment perpendicular to the 2d plane. The crystal lattice with lattice constant $a$ is  spanned by the  vectors ${\mathbf a}_1=a(\sqrt 3/2,1/2)$ and 
${\mathbf a}_2=a(-\sqrt 3/2,1/2)$ corresponding to the reciprocal vectors ${\mathbf b}_1=2\pi a^{-1}(1/\sqrt 3,1)$ and ${\mathbf b}_2=2\pi a^{-1}(-1/\sqrt 3,1)$.
The reciprocal lattice with the irreducible Brillouin zone is shown in Fig.~\ref{LatticeFig}(left). 
The interaction between two particles separated by a distance $r$ in the plane is
\begin{equation}
U(r)=\frac{D^2}{r^3}\text{ dipoles}\hspace{0.5cm}U(r)=\frac{Q^2}{r}\text{ charges}
\end{equation}
where $D^2=d^2/4\pi\epsilon_0$ for electric dipoles with dipole moment $d$, and $Q^2=q^2/4\pi\epsilon_0$ for particles with charge $q$.
We set  $\hbar=k_B=1$ in the following.

\subsection{Classical elasticity of point dipoles and point charges}
We  find the phonon modes of the  potential energy in the harmonic approximation in the usual way. To accelerate 
the convergence of the  sums, we use the Ewald summation technique as detailed  in Appendix \ref{Ewald}. 
In Fig.~\ref{LatticeFig} we plot the resulting two phonon branches along the vector ${\mathbf b}_1$ for the dipoles (middle) and the 
charged particles (right). 
%%%%%%%%%%%%%%%%%%%%%%%%%%%%%%%%%%%%%%%%%%%%%%%%%%%%
\begin{figure}
\epsfig{file=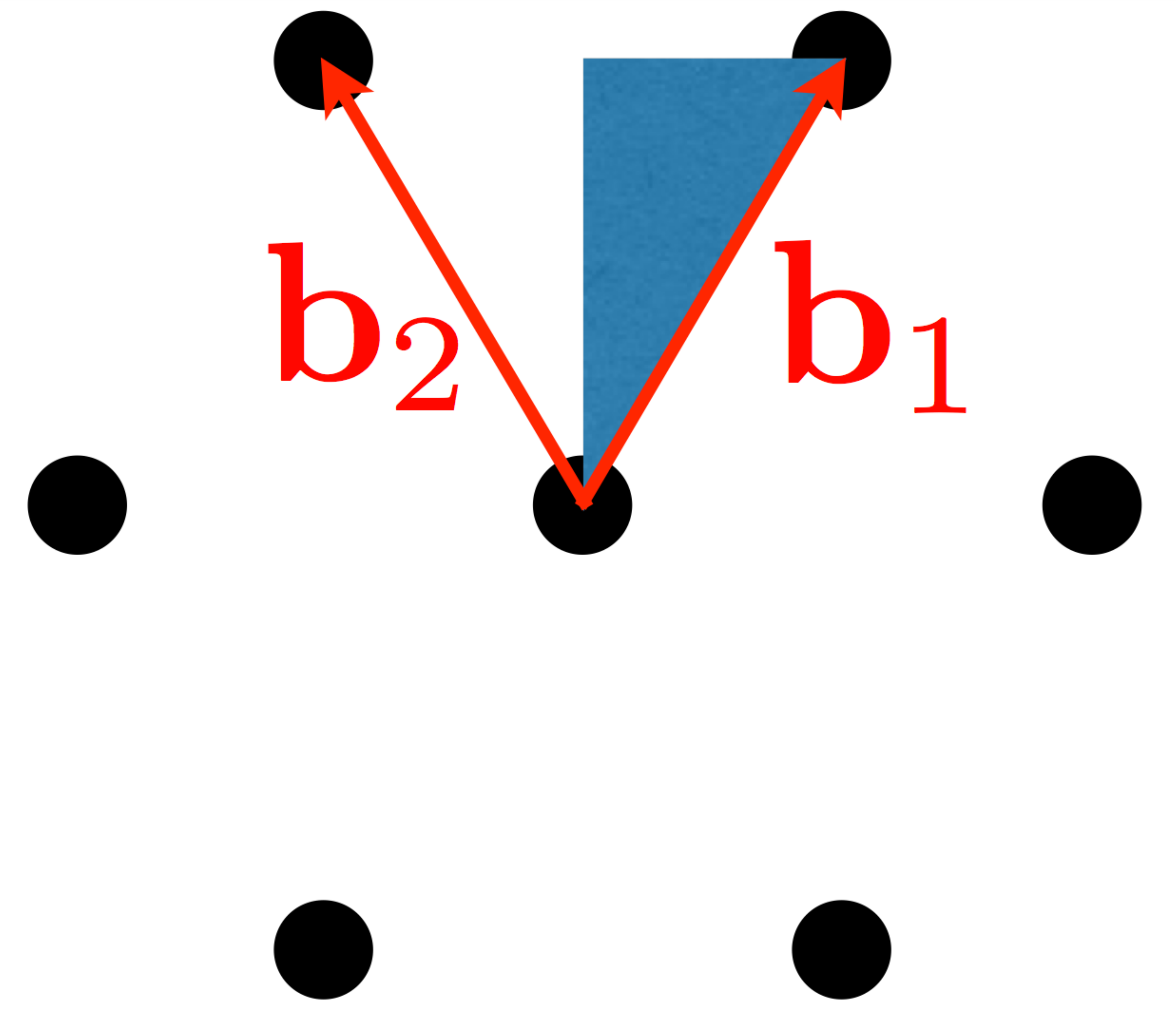,width=0.17\columnwidth}
\epsfig{file=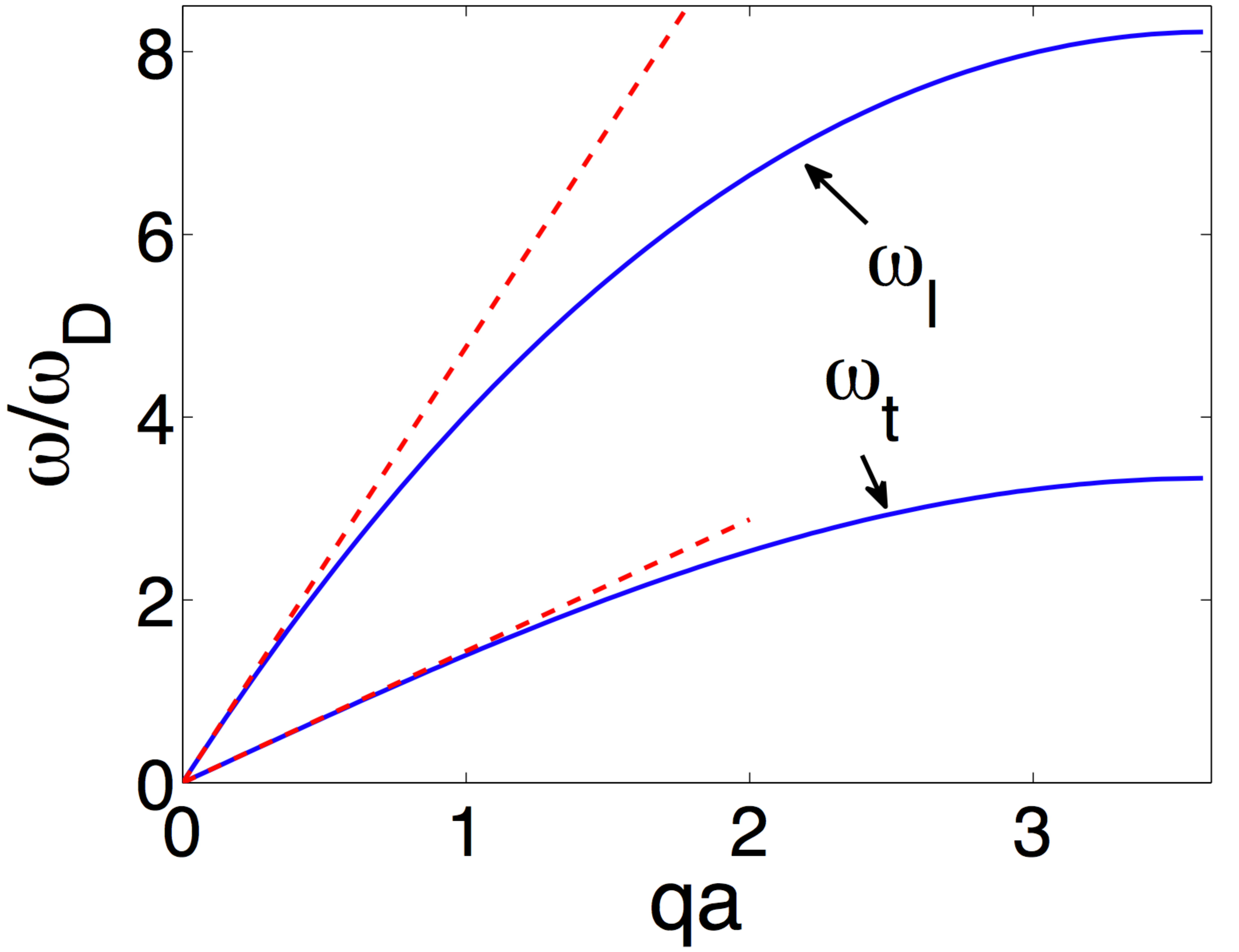,width=0.4\columnwidth}
\epsfig{file=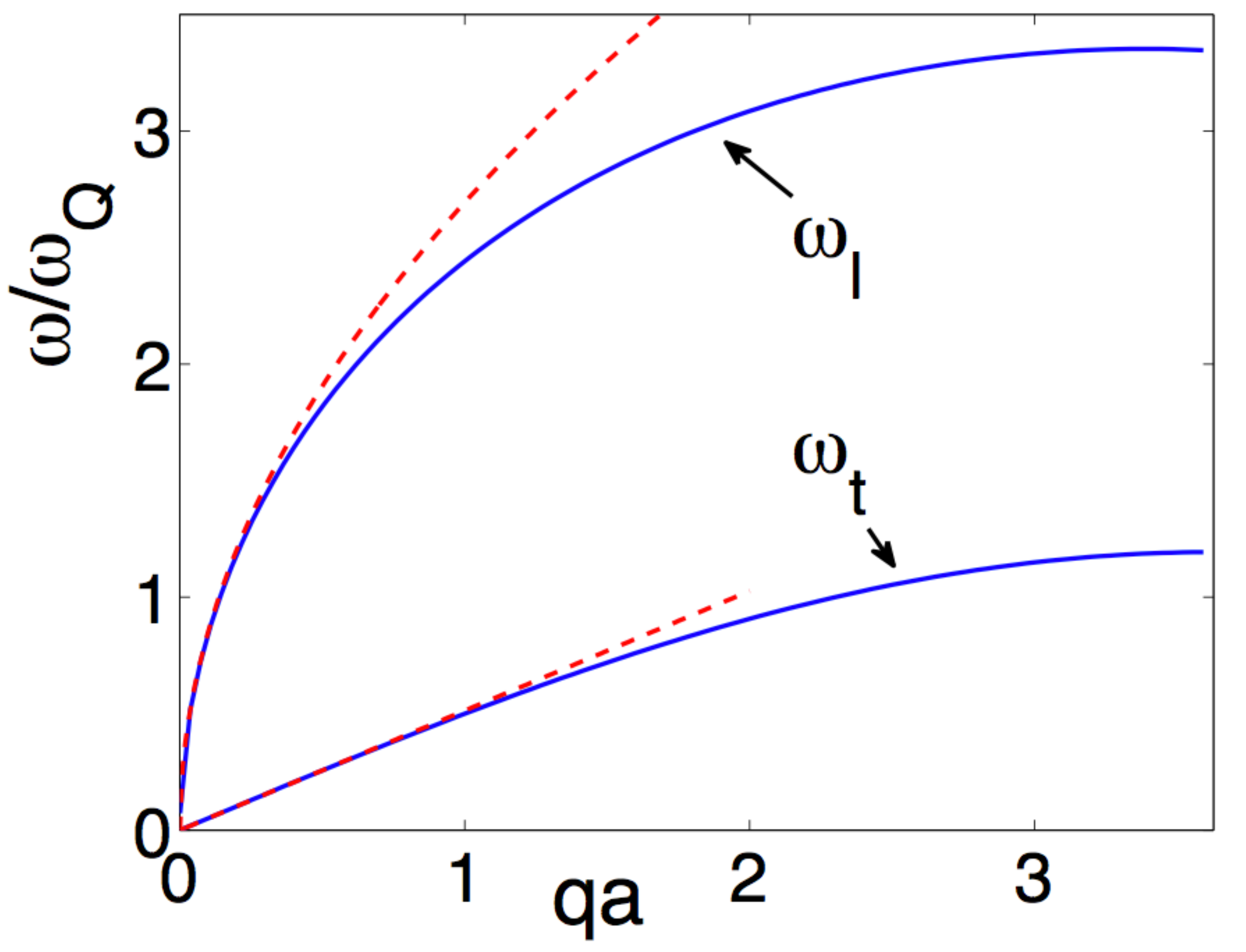,width=0.4\columnwidth}
\caption{Left: The hexagonal lattice in reciprocal space and the irreducible first Brillouin zone. Middle: The two phonon modes along  ${\mathbf b}_1$ for dipoles. Right: 
The two phonon modes along ${\mathbf b}_1$ for charged particles. Note that $\omega_l(q)\sim \sqrt q$ for small $q$, indicating an infinite bulk modulus in the 
long wavelength limit. The dashed lines are from Eqs.~(\ref{SoundDipoles})-(\ref{SoundCoulomb}).}
\label{LatticeFig}
\end{figure}
%%%%%%%%%%%%%%%%%%%%%%%%%%%%%%%%%%%%%%%%%%%%%%%%%%%%
The low energy mode  is purely transverse and the high energy mode is purely longitudinal for  long wavelengths 
where the hexagonal crystal is equivalent to an isotropic continuum system.~\cite{LandauLifshitz}
The characteristic phonon energy for the dipolar crystal is $\omega_D=\sqrt{D^2/ma^5}$, and a crystal of  Coulomb charges,  it is $\omega_C=\sqrt{Q^2/ma^3}$ 
where $m$ is the particle mass. 
For the dipoles, we find for long wave lengths the isotropic modes 
\begin{equation}
\omega_l(q)\simeq4.8\omega_D qa \hspace{0.4cm}\text{and}\hspace{0.4cm}\omega_t(q)\simeq1.4\omega_Dqa.
\label{SoundDipoles}
\end{equation}
These sound velocities differ somewhat from what was reported recently,~\cite{Lu} but as we shall demonstrate shortly, they accurately recover 
 well established values for the 
$k=0$ Lam\'e coefficients~\cite{Grunberg} which gives us confidence in our numerical calculations. 
For the charged particles, we have  plotted the long wave length formulas, 
\begin{equation}
\omega_l(q)=\frac{2\sqrt{\pi}}{3^{1/4}}\omega_C\sqrt{qa}\hspace{0.3cm}\text{and}\hspace{0.3cm}\omega_t(q)=\frac{2^{1/4}\eta^{1/2}}{3^{1/8}}\omega_Cqa
\label{SoundCoulomb}
\end{equation}
for the longitudinal and transverse mode respectively with $\eta=0.25$.~\cite{Fisher} We see that the numerics reproduce these  results confirming that the 
Ewald summation has converged. Note that the longitudinal mode scales as $\sqrt q$ for small momenta reflecting the long range nature of the Coulomb interaction.

The Lam\'e coefficients are defined by writing the  elastic energy of the crystal as~\cite{ChaikinBook} 
\begin{equation}
F_{\rm el}=\frac 1 {2L^2}\sum_{\mathbf k}\left\{\mu({\mathbf k})| u_t({\mathbf k})|^2
+[2\mu({\mathbf k})+\lambda({\mathbf k})]| u_l({\mathbf k})|^2\right\}k^2
\label{Fel}
\end{equation}
with $|{\mathbf u}({\mathbf k})|^2={\mathbf u}({\mathbf k}){\mathbf u}(-{\mathbf k})$ and $L^2$ the area of the system. 
  The longitudinal component of the displacement field is $u_l$, and
 $u_t$ is the transverse component. Note that "transverse" and "longitudinal"
 simply refers to the lowest and highest phonon mode for a given ${\mathbf k}$, since the eigenvectors are not in general parallel or 
 perpendicular to ${\mathbf k}$ when lattice effects are taken into account. The relation between the Lam\'e coefficients 
 and the phonon modes is then as usual 
\begin{equation}
\omega_t({\mathbf k})=\sqrt{\frac{\mu({\mathbf k})}{\rho}}k\hspace{0.5cm}\text{and}\hspace{0.5cm}\omega_l({\mathbf k})=\sqrt{\frac{2\mu({\mathbf k})+\lambda({\mathbf k})}{\rho}}k
\label{Lame}
\end{equation}
where  $\rho=m2/\sqrt3a^2$ is the mass areal density. 
 The natural scale for the Lam\'e coefficients is $D^2/a^5$ for dipoles and $Q^2/a^3$ for charged particles, 
 and they are  ${\mathbf k}$ dependent  due to lattice effects. Since $\omega_l({\mathbf k})\propto \sqrt k$ for $k\rightarrow 0$, 
 the Lam\'e coefficient  $\lambda({\mathbf k})$ diverges 
as $1/\sqrt k$ for the charged particles. 

\begin{figure}
\epsfig{file=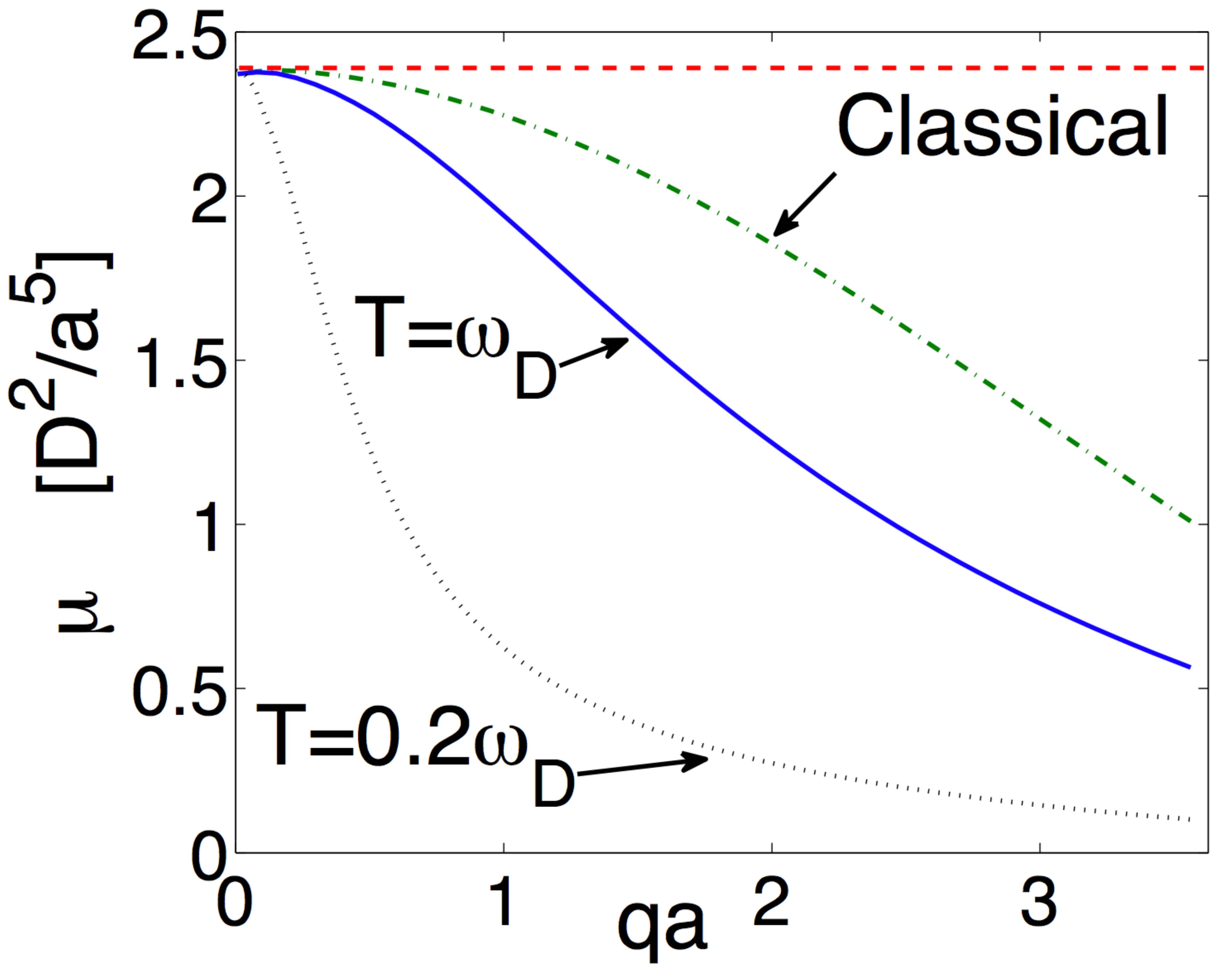,width=0.49\columnwidth}
\epsfig{file=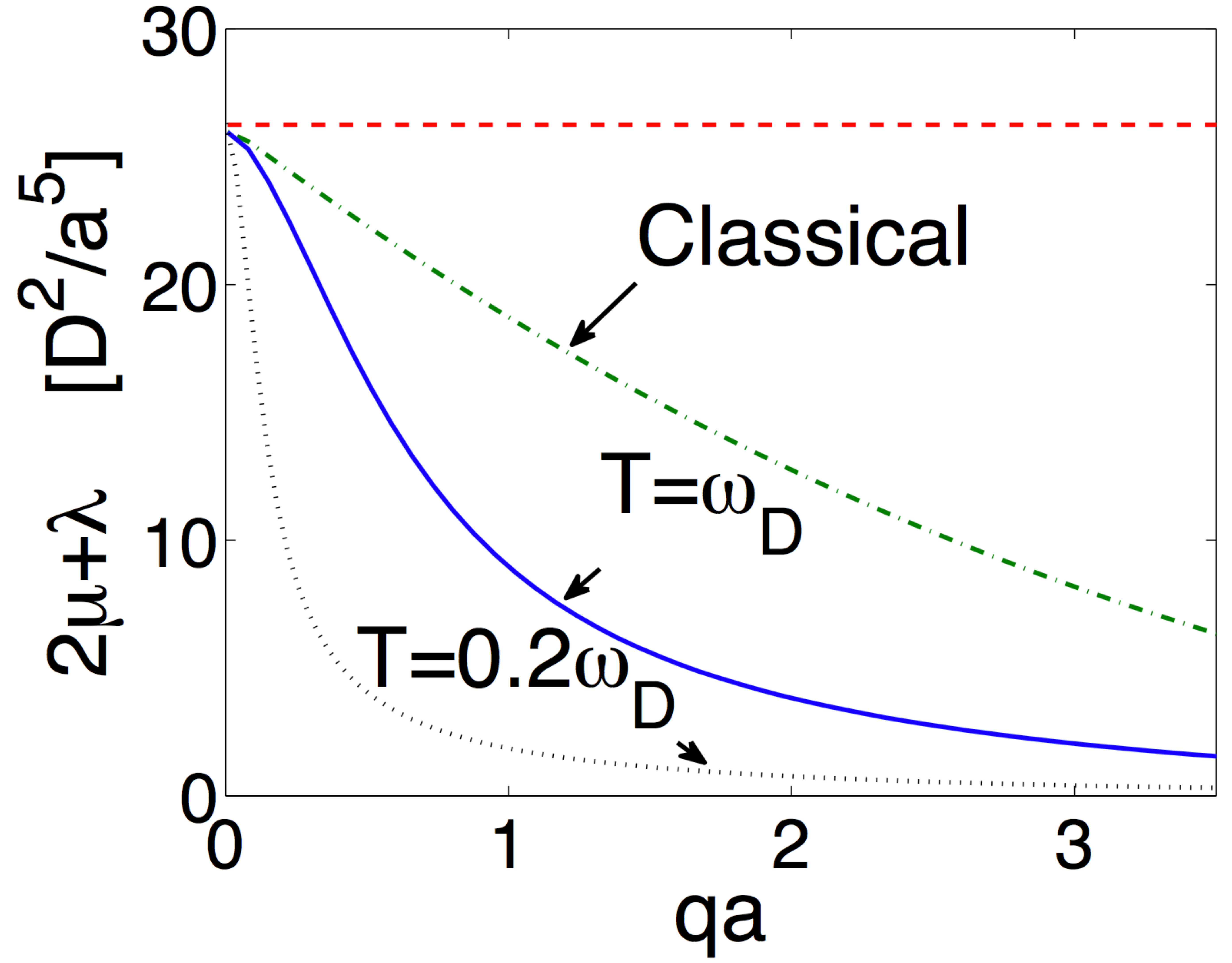,width=0.49\columnwidth}
\caption{Left: The Lam\'e coefficient for the lowest (transverse) mode of the  dipoles along ${\mathbf b}_1$. The solid and dotted lines include quantum effects for 
two different temperatures, and the dash-dotted line gives the classical limit. 
Right: The same for the highest (longitudinal) mode of the dipolar system. 
The dashed lines are the $q=0$ results  from  Eq.~(\ref{LameDipoles}).
Note the significant downward renormalization due to quantum fluctuations.
}
\label{LameFigDipoles}
\end{figure}

\begin{figure}
\epsfig{file=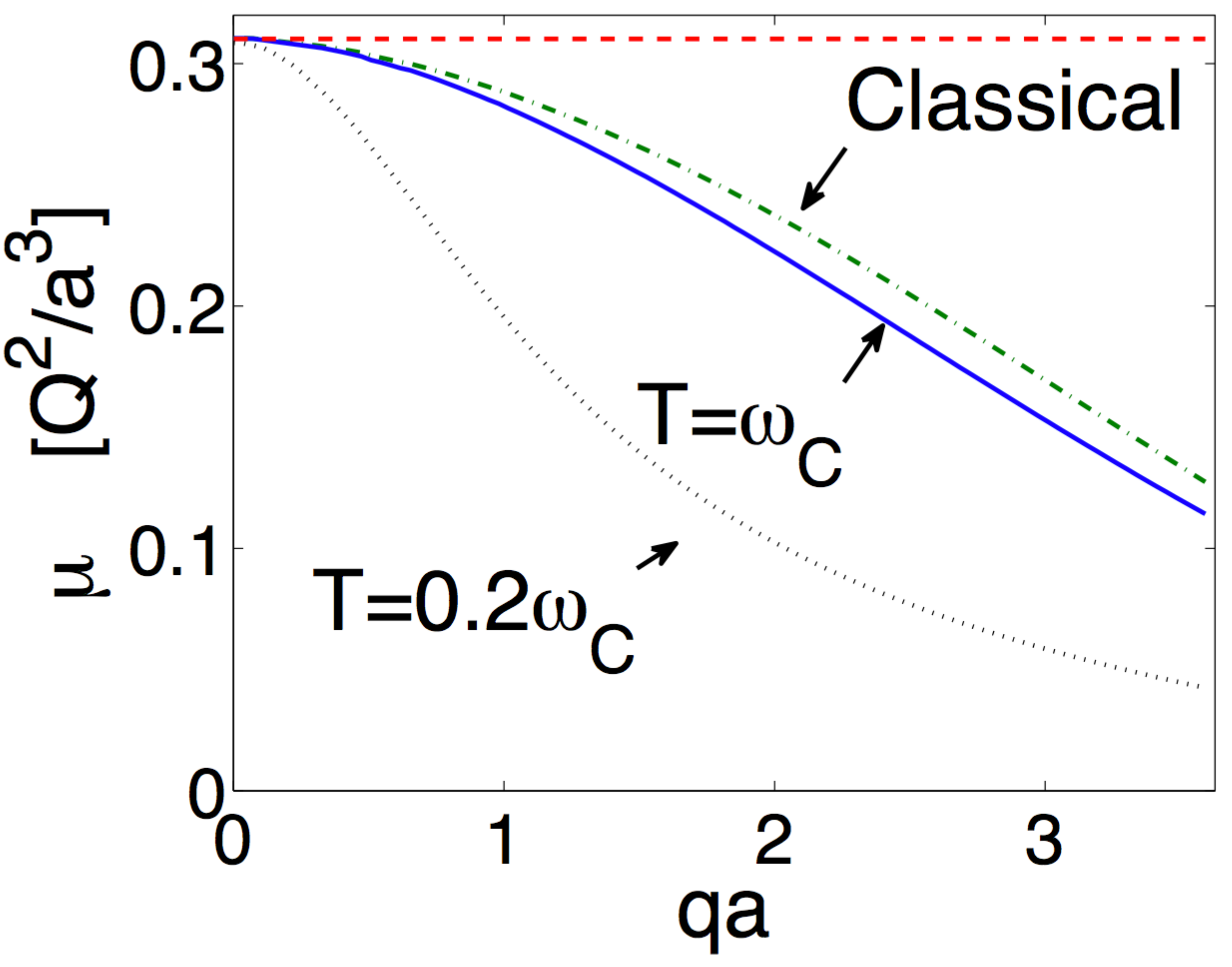,width=0.49\columnwidth}
\epsfig{file=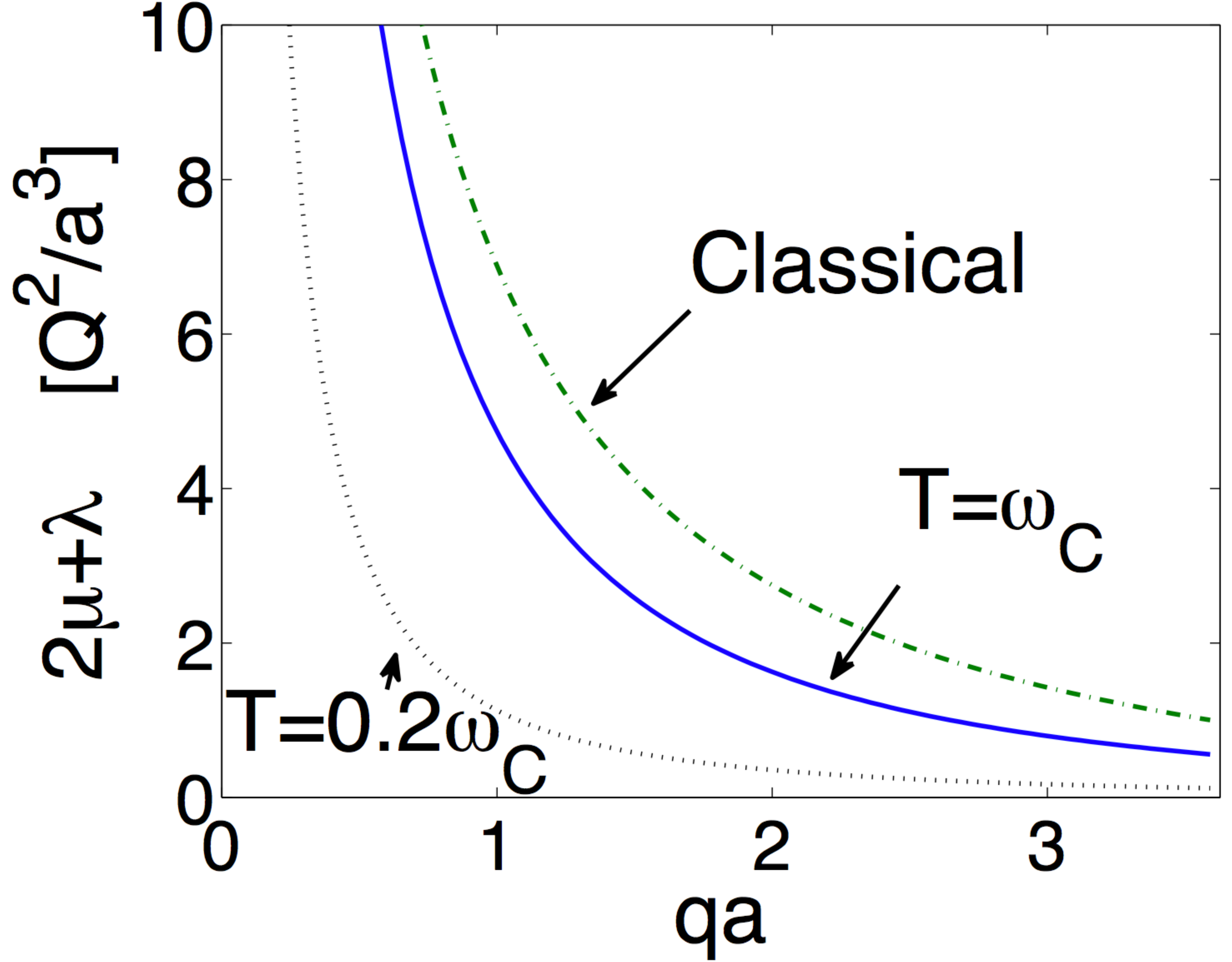,width=0.49\columnwidth}
\caption{Left: The Lam\'e coefficient for the lowest (transverse) mode of the  charges particles along ${\mathbf b}_1$. The solid and dotted lines include quantum effects for 
two different temperatures, and the dash-dotted line gives the classical limit. 
  Right: The same for the highest (longitudinal) mode of the charged particles. The dashed line is obtained from  Eq.~(\ref{LameCharges}).
}
\label{LameFigCoulomb}
\end{figure}

In Figs.~\ref{LameFigDipoles}-\ref{LameFigCoulomb}, we plot the classical Lam\'e coefficients along ${\mathbf b}_1$  for the dipoles and 
the charged particles. The elastic parameters display a  significant $k$-dependent  softening due to the discrete lattice symmetry.
We also plot in Fig.~\ref{LameFigDipoles}  the $k=0$ Lam\'e coefficients corresponding to Eq.~(\ref{SoundDipoles}), i.e.
\begin{equation}
\mu({0})\simeq2.4\frac{D^2}{a^5}\hspace{0.3cm}\text{and}\hspace{0.3cm}2\mu({0})+\lambda({0})\simeq26\frac{D^2}{a^5}.
\label{LameDipoles}
\end{equation}
These values agree very well those reported in Ref.~\onlinecite{Grunberg}.
Likewise, we plot in Fig.~\ref{LameFigCoulomb} the  $k=0$ value for the transverse mode corresponding to Eq.~(\ref{SoundCoulomb}), i.e.~\cite{Fisher}  
\begin{equation}
\mu({0})=\eta\frac{2^{3/2}Q^2}{3^{3/4}a^3}
\label{LameCharges}
\end{equation}
which is recovered by our numerics. 
%%%%%%%%%%%%%%%%%%%%%%%%%%%%%%%%%%%%%%%%%%%%%%%%%%%%
%%%%%%%%%%%%%%%%%%%%%%%%%%%%%%%%%%%%%%%%%%%%%%%%%%%%
%%%%%%%%%%%%%%%%%%%%%%%%%%%%%%%%%%%%%%%%%%%%%%%%%%%%
%%%%%%%%%%%%%%%%%%%%%%%%%%%%%%%%%%%%%%%%%%%%%%%%%%%%

Using Eq.\ (\ref{Fel}), it is straightforward to calculate the mean displacement of the particles from their equilibrium positions at a given temperature $T$, and we obtain 
\begin{eqnarray}
\langle u_l({\mathbf k})u_l({\mathbf k}')\rangle&=&\delta_{{\mathbf k},-{\mathbf k}'}L^2\frac{T}{[2\mu({\mathbf k})+\lambda({\mathbf k})]k^2}\nonumber\\
\langle u_t({\mathbf k})u_t({\mathbf k}')\rangle&=&\delta_{{\mathbf k},-{\mathbf k}'}L^2\frac{T}{\mu({\mathbf k})k^2}.
\label{uuk}
\end{eqnarray}

\subsection{Quantum softening of the Lam\'e coefficients}
We now include quantum effects on the Lam\'e coefficients by quantizing the phonons. This can be done in several ways. Here, we use the path 
integral approach since it allows us to describe the quantum effects on the crystal melting in terms of a simple geometrical picture of wiggling 
particle trajectories described in the introduction. 
For completeness, we also present a canonical quantization approach   in  Appendix \ref{Canonical}. 

The partition function $Z$ for the crystal can be written as an integral over all possible paths ${\mathbf u}({\mathbf r},\tau)$
of the particle displacements in imaginary time $\tau$  as~\cite{FeynmanHibbs} 
\begin{equation}
Z=\int {\mathcal D}[{\mathbf u}({\mathbf k},\tau)]e^{-\int_0^\beta d\tau[F_{\rm kin}(\tau)+F_{\rm el}(\tau)]}
\label{Zpath}
\end{equation}
where $F_{\rm kin}(\tau)=\rho \sum_{\mathbf k}|\partial_\tau{\mathbf u}({\mathbf k},\tau)|^2/2L^2$ is the kinetic energy, $\rho$ is the 2d mass density,  and the elastic energy 
$F_{\rm el}(\tau)$ is given by Eq.\ (\ref{Fel}) with the replacement ${\mathbf k}\rightarrow ({\mathbf k},\tau)$. 
We do  not include permutations of the particle positions  at the boundary $\tau=\beta$ of the imaginery time slab, 
so the boundary condition in Eq.\ (\ref{Zpath}) is  ${\mathbf u}({\mathbf r},\beta)={\mathbf u}({\mathbf r},0)$. At this level 
of approximation, we therefore cannot distinguish between bosonic and fermionic particles.  We can from Eq.\ (\ref{Zpath}) calculate the 
fluctuations of the particles including quantum effects, 
\begin{gather}
\langle  u_\sigma({\mathbf k},\tau) u_\sigma(-{\mathbf k},\tau)\rangle=
\frac{L^2\beta}{\rho}\sum_n\frac1{\omega_n^2+\omega_\sigma({\mathbf k})^2}\nonumber\\
=\frac{L^2}{2\rho\omega_\sigma({\mathbf k})}\coth[\beta\omega_\sigma({\mathbf k})/2)]
\label{Quantumuu}
\end{gather}
where the sum is over Matsubara frequencies $\omega_n=2n\pi T$ and phonon modes $\sigma=t,l$. Recasting this result in the form of Eq.~(\ref{uuk})
defines the quantum Lam\'e coefficients $\mu_Q({\mathbf k})$ and $\lambda_Q({\mathbf k})$ as 
\begin{gather}
\mu_Q({\mathbf k})=\frac{2T\sqrt{\rho\mu}}{k}\tanh\left(\sqrt{\frac{\mu}{\rho}}\frac{k}{2T}\right)\nonumber\\
2\mu_Q({\mathbf k})+\lambda_Q({\mathbf k})=\frac{2T\sqrt{\rho(2\mu+\lambda)}}{k}\tanh\left(\sqrt{\frac{2\mu+\lambda}{\rho}}\frac{k}{2T}\right),
\label{LameQuantum}
\end{gather}
where we have suppressed the $k$-dependence of the classical Lam\'e coefficients $\mu({\mathbf k})$ and  $\lambda({\mathbf k})$ for notational simplicity.  
Equation (\ref{LameQuantum}) reveals  that the magnitude of quantum effects on the Lam\'e coefficients is determined by the parameters 
$\sqrt{\mu/\rho}k/T=\omega_t/T$ and $\sqrt{(2\mu+\lambda)/\rho}k/T=\omega_l/T$.
For $\omega_\sigma/T\rightarrow 0$ we recover the classical results given by Eq.\ (\ref{Lame}), whereas the elastic coefficients are 
decreased due to quantum fluctuations whenever  $\omega_\sigma/T\gtrsim 1$.
 
In Figs.~\ref{LameFigDipoles}-\ref{LameFigCoulomb}, we plot the quantum Lam\'e coefficients  along ${\mathbf b}_1$ 
for $T/\omega_D=T/\omega_C=1$ and $T/\omega_D=T/\omega_C=0.2$. 
We see that quantum effects significantly soften the crystal for decreasing temperature, and that it is 
the high energy fluctuations which are reduced the  most. Quantum softening is therefore greater for the longitudinal mode,
reducing $[2\mu_Q({\mathbf k})+\lambda_Q({\mathbf k})]/[2\mu({\mathbf k})+\lambda({\mathbf k})]$ more than  $\mu_Q({\mathbf k})/\mu({\mathbf k})$.

\section{Modified Lindemann criteria for crystal and hexatic melting}
Since there is algebraic, as opposed to long range translational order in a  2d crystal when  $T>0$,~\cite{MerminWagner} it is not possible to estimate the melting 
temperature of the crystal from a usual Lindemann criterion, as discussed in the introduction. 
We will therefore use a modified Lindemann criterium for the melting. 
Our basic assumption is  that the melting of the crystal occurs in two steps with increasing temperature.~\cite{HalperinNelson,NelsonHalperin,NelsonBook} 
First, the crystal melts at a temperature $T_m$ into a hexatic phase,  characterised by long range bond angle order but short range translational order. 
Then, at a higher temperature $T_i$ the hexatic phase melts  into an isotropic liquid. The existence of the hexatic phase is well established for 
classical systems~\cite{Gasser},  but our knowledge concerning its stability against quantum fluctuations is limited. A Monte-Carlo study 
supports the existence of such a phase in the quantum regime in the case of distinguishable  particles with Coulomb interactions.~\cite{Clark} We therefore 
focus on how quantum fluctuations affect on the stability of the hexatic phase.

\subsection{Melting of the crystal phase}
To calculate the  temperature $T_m$ where the crystal melts into the hexatic phase, 
we will use the modified Lindemann criterion given by Eq.\ (\ref{Lindemann}). It states that  the crystal melts when 
the \emph{relative} fluctuations of the particle positions of two nearest neighbours are larger than the lattice constant $a$. 
Using the quantum Lam\'e coefficients in Eq.~(\ref{uuk}) yields 
\begin{gather}
\delta u_i^2=\langle |u_i({\mathbf r})-u_i(0)|^2\rangle=T\frac2{L^2}\sum_{\mathbf k}(1-\cos{\mathbf k}\cdot{\mathbf r})\nonumber\\
\times\left[\frac{\epsilon_{l,i}(\mathbf{k})^2}{[2\mu_Q({\mathbf k})+\lambda_Q({\mathbf k})]k^2}+
\frac{1-\epsilon_{l,i}(\mathbf{k})^2}{\mu_Q({\mathbf k})k^2}\right]
\label{uur}
\end{gather}
for the fluctuations along a specific direction $i=x$ or $i=y$. Here, 
$\epsilon_{l,i}(\mathbf{k})$ is $i$'th component of the eigenvector of the highest mode.

The melting temperature for the crystal phase in the classical limit was reported  to be
$T_m\simeq 0.0907{D^2}/{a^3}$ for dipoles,~\cite{Kalia,Grunberg} and $T_m\simeq 0.0136{Q^2}/{a}$ for charged particles.~\cite{Muto}
From these results, we can determine the Lindemann number in the classical regime. Using Eqs.~(\ref{Lindemann}) and (\ref{uur}) with the classical Lam\'e coefficients
to calculate the particle fluctuations at the classical melting temperature yields 
\begin{equation}
\gamma_{m,cl}=0.14\text{ dipoles}\hspace{0.5cm}\gamma_{m,cl}=0.15\text{ charges}.
\label{LindemannClass}
\end{equation}
We see that the classical Lindemann numbers are essentially  the same for the dipoles  and the charged particles. This result suggests that 
  melting is mainly determined by the geometry of the crystal,
 depending very weakly on the detailed form of the interaction potential so that the Lindemann number is almost universal.

As discussed in the introduction, we expect that the Lindemann numbers in general depend on temperature due to quantum effects. 
We can  determine the Lindemann number $\gamma_m$  at  $T=0$ using recent quantum Monte-Carlo results
suggesting  a $T=0$ quantum phase transition between the crystal phase and the liquid phase: 
For dipoles, two independent calculations give the value 
$r_D\simeq 18\pm 4$ ~\cite{Astrakharchik,Buchler},  and for distinguishable charged particles one obtains the value $r_C\simeq 127$ 
for the critical value of this quantum phase transition.~\cite{Clark}  Here, the $r$-parameters 
\begin{equation}
r_D=\frac{mD^2}{a}\text{ dipoles}\hspace{0.5cm}r_C=mQ^2a\text{ charges}
\end{equation}
are the ratios between the nearest neighbour interaction energy and the quantum kinetic energy. 
Using the $T\rightarrow 0$ limit of Eq.~(\ref{LameQuantum}) in Eq.~(\ref{uur}), we obtain  
\begin{gather}
\langle |u_i({\mathbf r})-u_i(0)|^2\rangle=\frac1{\rho L^2}\sum_{\mathbf k}(1-\cos{\mathbf k}\cdot{\mathbf r})\nonumber\\
\times\left[\frac{\epsilon_{l,i}(\mathbf{k})^2}{\omega_l({\mathbf k})}+
\frac{1-\epsilon_{l,i}(\mathbf{k})^2}{\omega_t({\mathbf k})}\right]
\label{uurQ}
\end{gather}
for $T=0$. Equation (\ref{uurQ}) predicts as expected
 that the zero point motion of the particles  scale as the typical  harmonic oscillator length  for the phonons, i.e.\ $\delta u^2\sim 1/m\omega_D$ 
for dipoles and $\delta u^2\sim 1/m\omega_C$ for charged particles. Using Eq.\ (\ref{uurQ}), 
we can determine the Lindemann number for $T=0$ at the critical coupling strength  for the quantum melting transition, obtaining 
\begin{equation}
\gamma_{m,0}=0.31\text{ dipoles}\hspace{0.5cm}\gamma_{m,0}=0.31\text{ charges}.
\label{LindemannT0}
\end{equation}
 The Lindemann numbers are again the same for the dipoles and the charges at this level of accuracy, indicating that 
the melting of the crystal phase is primarily  determined by geometry also at $T=0$.
The $T=0$ value of $\gamma_m$ at the critical point is consistent with what was obtained using perturbation theory.~\cite{Lozovik2}

Comparing  Eqs.~(\ref{LindemannClass}) and (\ref{LindemannT0})
shows that the Lindemann numbers are significantly larger at $T=0$ than in the classical regime. As discussed in the introduction, 
the path integral approach provides a simple geometrical interpretation of this result.  
Indeed, from Eq.~(\ref{Zpath}) it is clear that the quantum problem corresponds to the  melting of a crystal of lines in a 3d slab of thickness $\beta$, see Fig. \ref{WorldlineFig}. 
Only when $\beta\rightarrow 0$ does one recover the classical problem of the melting of a 2d crystal.
Since the lines can wiggle significantly along the $\beta$-direction without melting the crystal, it is natural to expect that the Lindemann 
number is larger in the quantum regime  as compared to the classical regime.

\subsection{Melting of the hexatic phase}
In the two-step melting scenario, the system is in a hexatic phase characterised by extended bond angle order for temperatures $T_m<T<T_i$. 
We  therefore use the Lindemann criterion based on the fluctuations in  the bond angle $\theta$ given by Eq.\ (\ref{LindemannHexatic}).
 Fourier transforming  Eq.\ (\ref{uuk}) using the quantum Lam\'e coefficients gives after some algebra 
\begin{equation}
\langle\theta^2\rangle=\frac T{4L^2}\sum_{\mathbf k}\left[\frac{(\mathbf{k}\times\mathbf{\epsilon}_l(\mathbf{k}))^2}{[2\mu_Q({\mathbf k})+\lambda_Q({\mathbf k})]k^2}+
\frac{(\mathbf{k}\times\mathbf{\epsilon}_t(\mathbf{k}))^2}{\mu_Q({\mathbf k})k^2}\right]
\label{AngleFluc}
\end{equation}
where $\epsilon_t(\mathbf{k})$ is the eigenvector of the lowest mode. 
For an isotropic medium where $\epsilon_l\parallel{\mathbf k}$  and $\epsilon_t\perp{\mathbf k}$, Eq.\ (\ref{AngleFluc}) reduces to 
\begin{equation}
\delta\theta^2=\frac T{4L^2}\sum_{\mathbf k}\frac 1{\mu_Q({\mathbf k})},
\label{AngleFlucSimple}
\end{equation}
i.e.\ the bond angle fluctuations are determined by the transverse mode only. Since a 
hexagonal crystal is equivalent to an isotropic medium for long wave lengths,\cite{LandauLifshitz} and since it is these low energy modes which contribute most to the 
fluctuations, Eq.~(\ref{AngleFlucSimple}) turns out to be a very good approximation to Eq.~(\ref{AngleFluc}).

To determine the Lindemann numbers for the hexatic phase, we again use results for the melting temperatures reported in the literature. 
The melting temperature of the hexatic phase in the classical limit was found to be $T_i\simeq 0.0968U(a)$ with $U(a)=D^2/a^3$ for  dipoles~\cite{Grunberg}, and 
$T_i\simeq 0.0159U(a)$ with $U(a)=Q^2/a$ for  charged particles.~\cite{Muto} Using these temperatures and the classical Lam\'e coefficients in  Eq.~(\ref{AngleFluc}) yields
\begin{equation}
\gamma_{i,cl}=0.12\text{ dipoles}\hspace{0.5cm}\gamma_{i,cl}=0.13\text{ charges}
\label{LindemannHexaticClass}
\end{equation}
for the Lindemann numbers determining the melting of the hexatic phase  in the classical regime. As for the crystal phase, the Lindemann numbers are essentially the 
same for the dipoles and the charged particles, suggesting again  that melting of the hexatic phase is a geometric phenomenon, largely independent of the 
precise form of the interaction.

The $T=0$ Monte-Carlo calculations for the dipoles did not examine the quantum hexatic phase,~\cite{Astrakharchik,Buchler} so it is presently 
not known whether it exists all the way down to $T=0$. In the case of distinguishable "Boltzmannian" charged particles, it was found that the hexatic phase persists 
to quite low temperatures where quantum effects are significant, disappearing in a tricritical point at $T\simeq 0.04U(a)$.~\cite{Clark} 
Since our analysis indicates that, for a given value of $n_0\Lambda_T^2$, the melting is insensitive to  the detailed form of the interaction potential, 
this Monte-Carlo result suggests that the hexatic phase persist deep into the low temperature regime both for dipoles and 
for charged particles. It is therefore interesting to evaluate the Lindemann numbers at $T=0$ for the bond angle fluctuations. Using the  $T=0$ limit of Eq.~(\ref{AngleFluc}), 
\begin{equation}
\delta\theta^2=\frac 1{8\rho L^2}\sum_{\mathbf k}\left[\frac{(\mathbf{k}\times\mathbf{\epsilon}_l(\mathbf{k}))^2}{\omega_l({\mathbf k})}+
\frac{(\mathbf{k}\times\mathbf{\epsilon}_t(\mathbf{k}))^2}{\omega_t({\mathbf k})}\right],
\label{AngleFlucQ}
\end{equation}
 yields 
\begin{equation}
\gamma_{i,0}=0.23\text{ dipoles}\hspace{0.5cm}\gamma_{i,0}=0.24\text{ charges}
\end{equation}
at the quantum quantum transition points $r_D\simeq 18$ and $r_C\simeq 127$ for dipoles and charged particles respectively. 
Again, the angle fluctuations differ very little between the 
dipolar and the charged systems. 
The precise values of the Lindemann numbers may, of course, differ at the exact boundaries of  the hexatic phase in the quantum regime, 
which are presently unknown.

As we discussed, the melting of the hexatic phase is determined almost exclusively by the lowest phonon mode in the sense that 
Eq.~(\ref{AngleFlucSimple}) is an excellent approximation to Eq.~(\ref{AngleFluc}). In addition, the  lowest phonon mode 
is less affected by quantum fluctuations than the highest phonon mode. It follows from this that
 the hexatic phase is more robust  towards quantum fluctuations compared 
to the crystal phase, and as a result the hexatic region in the phase diagram should \emph{increase} with decreasing potential/quantum-kinetic energy ratio $r$. 
 This conclusion depends of course on the assumption that the Lindemann numbers are 
independent of temperature, so we expect it to be valid only when the quantum corrections are small. 
When the temperature dependence of the Lindemann numbers is significant  we cannot determine the fate of the hexatic 
without further information. 

\section{Phase diagrams}
\begin{figure}
\epsfig{file=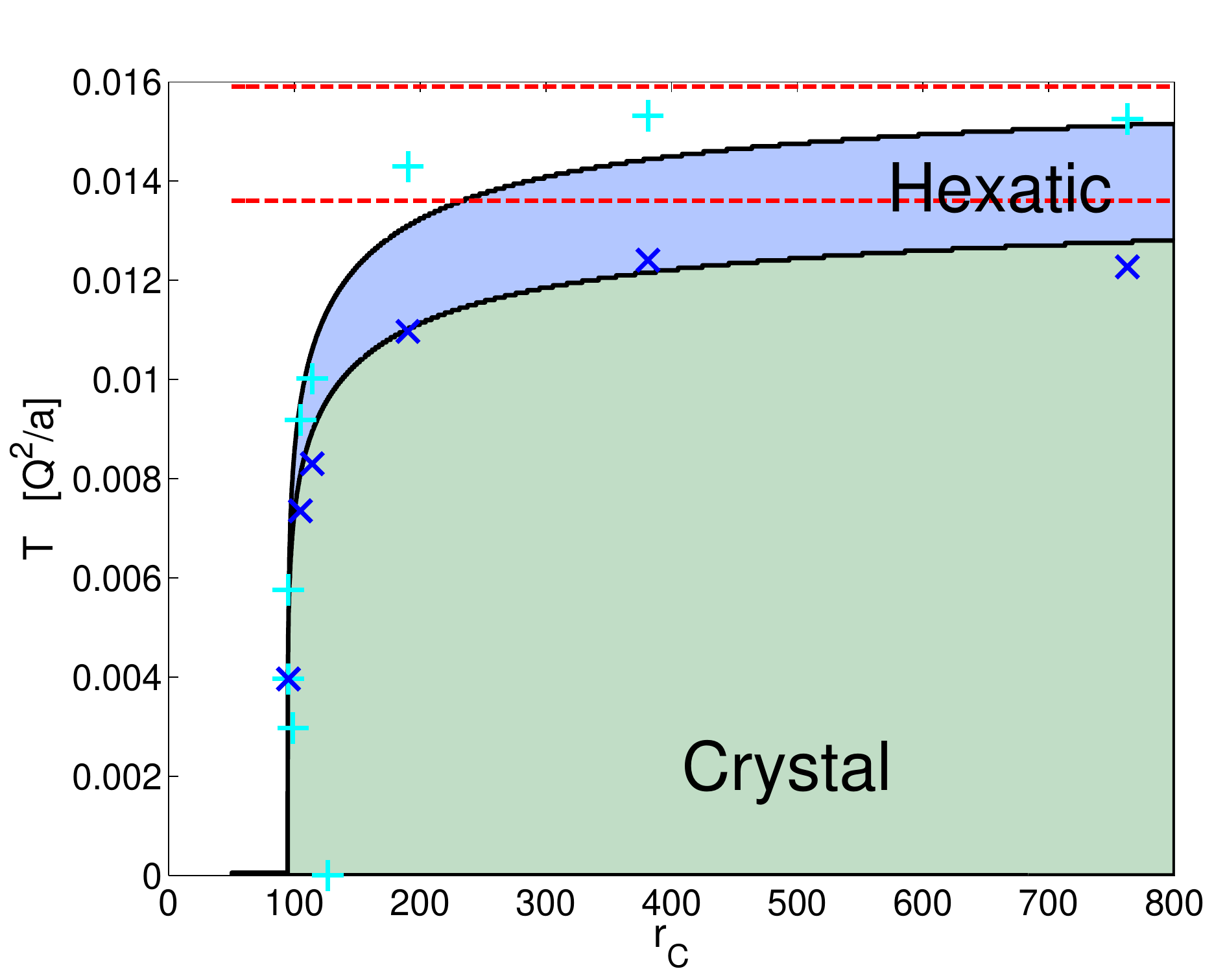,width=1\columnwidth}
\caption{Approximate phase diagram of the charged particles as a function of the rescaled temperature $T/(Q^2/a)$ and the 
potential/quantum-kinetic energy ratio $r_C$, 
using temperature dependent Lindemann constants determined by Eq.~(\ref{GammaT}) with 
$n=6$. The dashed lines are the classical limits for $r_C\gg 1$, and the $\times$'s and $+$'s are numerical Monte-Carlo data from Ref.~\onlinecite{Clark}. }
\label{CeperleyPhase}
\end{figure}

In this section, we provide approximate phase diagrams, using the fact that the melting of the crystal and 
hexatic phases are insensitive to  the detailed form of the interaction potential. 
However, by comparing the values of the Lindemann numbers in the classical regime and at $T=0$ given by
Eqs.~(\ref{LindemannClass}) and (\ref{LindemannT0}) we see that they depend  on temperature. 
To provide a tentative phase diagram, we therefore write the Lindemann number
determining the crystal melting  on the phenomenological form 
\begin{equation}
\gamma_m(T)=\gamma_{m,0}+(\gamma_{m,cl}-\gamma_{m,0})\left(\frac{T}{T_m}\right)^n,
\label{GammaT}
\end{equation}
which  interpolates between the $T=0$ value and the classical value for $T=T_m$.
We write $\gamma_i(T)$ in the same phenomenological form. To determine $n$, we compare the phase diagram produced  by
these phenomenological forms with the phase diagram obtained by Monte-Carlo calculations for distinguishable charged particles 
in Ref.~\onlinecite{Clark}. It turns out that $n=6$ yields a reasonable good fit as is shown in Fig.~\ref{CeperleyPhase}. It must be emphasised that 
we have not performed a systematic fit to determine the optimal value of $n$, since this is not relevant at this level of approximation, 
where our goal is simply to provide a qualitatively reliable phase diagram.  The detailed  form of the actual phase diagram could be 
different. For instance, our analysis does not reproduce the  tri-critical point  at $T\simeq0.004U(a)$ found in the Monte-Carlo calculations. Instead, we have chosen 
to let the hexatic and crystal phases continue down to $r_C\simeq95$ for $T=0$ corresponding to the Lindemann numbers $\gamma_{i,0}=0.25$ and $\gamma_{m,0}=0.33$.
Moreover, as stressed in the introduction, for the case of a Yukawa potential, a quantum hexatic phase could exist for a finite parameter range even at 
$T=0$.
%%%%%%%%%%%%%%%%%%%%%%%%%%%%%%%%%%%%%%%%%%%%%%%%%%%%
%%%%%%%%%%%%%%%%%%%%%%%%%%%%%%%%%%%%%%%%%%%%%%%%%%%%
Note that there is no significant increase in the temperature range for which the hexatic phase is stable with decreasing $r_C$ in Fig.~\ref{CeperleyPhase}. 
This is because the phenomenological form for the temperature dependence of the Lindemann numbers given by Eq.\ (\ref{GammaT}) with $n=6$
essentially cancels this effect   for the temperatures shown.

In Fig.~\ref{CeperleyPhase},  we also plot as dashed lines the critical temperatures in the classical limit. 
We see that the quantum suppression of the critical temperatures is significant even for $r_C\sim {\mathcal O}(100)$,
where one would naively expect the system to be well within the classical regime. The enhanced importance 
of quantum fluctuations arises because the classical melting temperatures of the crystal and hexatic phases are so low with $T_m/U(a)=0.0136\ll 1$ and 
$T_i/U(a)=0.0159\ll 1$. Since quantum softening of the Lam\'e coefficients sets in for
$T/\omega_C\lesssim 1$, and $T/\omega_C=\sqrt{r_C} T/U(a)$ with $T_m/U(a)\ll 1$ in the hexatic and crystal phases, 
the classical melting temperature is only recovered for $r_C\gg 1$.

\begin{figure}
\epsfig{file=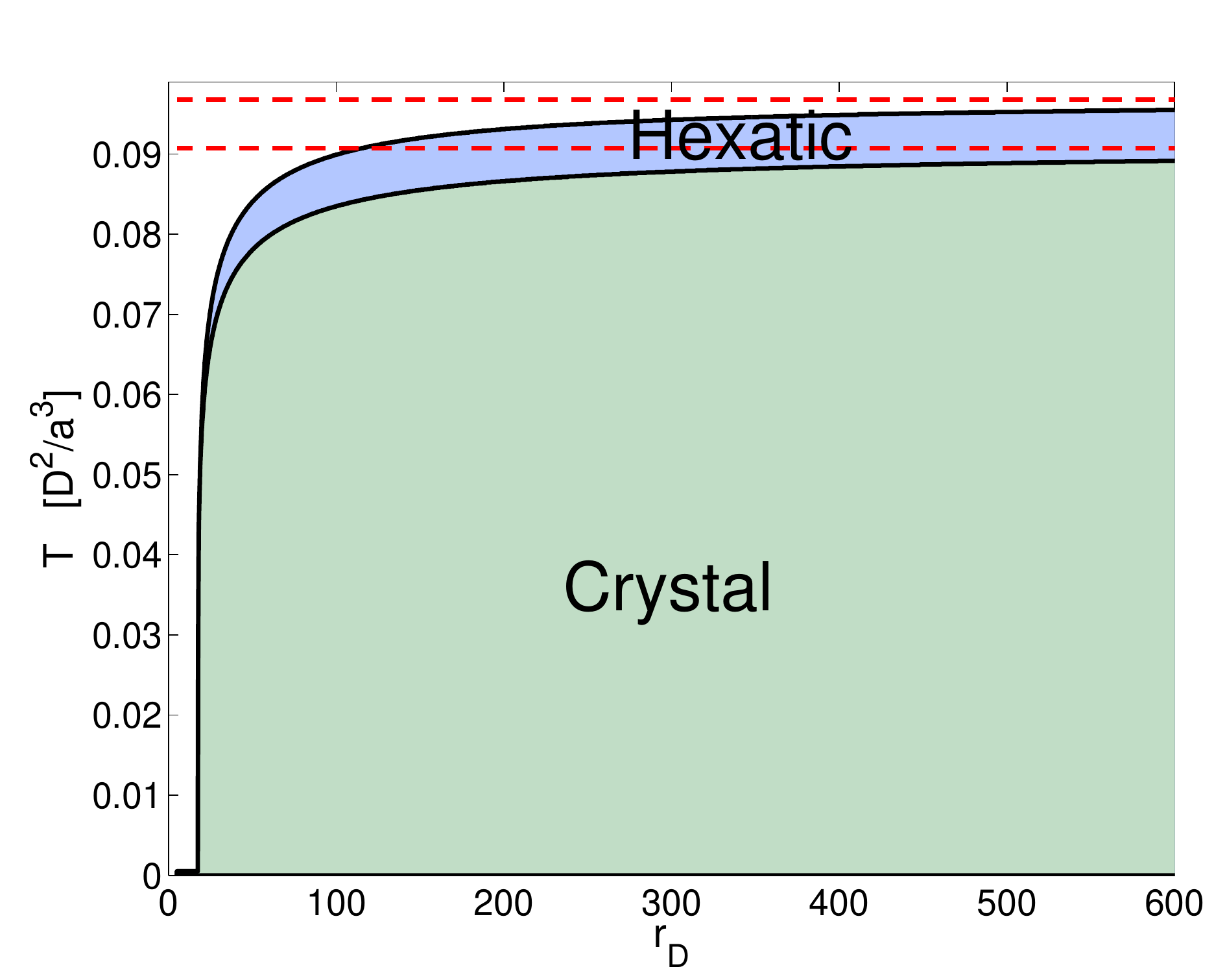,width=1\columnwidth}
\caption{Approximate phase diagram of the dipoles as a function of the rescaled temperature $T/(D^2/a^3)$ and the 
potential/quantum-kinetic energy ratio $r_D$, using temperature dependent Lindemann constants determined by Eq.~(\ref{GammaT}) with 
$n=6$. The dashed lines are the classical limits for $r_D\gg 1$. }
\label{DipolePhase}
\end{figure}

In Fig.~\ref{DipolePhase}, with applications to cold dipolar gases in mind,
we plot our tentative phase diagram for dipolar charges  obtained from the Lindemann criteria using 
temperature dependent Lindemann numbers given by 
Eq.~(\ref{GammaT}) with $n=6$. As for the charged case, 
 quantum effects on the melting are significant even for $r_D\gg 1$ since the classical critical temperatures 
are so low with $T_m/U(a)\ll 1$ and $T_i/U(a)\ll 1$. 
As for the case of charged particles, there is no significant increase in the temperature range where the hexatic phase is stable with decreasing $r_D$ due to the 
 chosen temperature dependence of the Lindemann numbers. 
Again, this phase diagram is only qualitatively reliable but it suggests that 
the hexatic phase extends well into the quantum regime.
%%%%%%%%%%%%%%%%%%%%%%%%%%%%%%%%%%%%%%%%%%%%%%%%%%%%
%%%%%%%%%%%%%%%%%%%%%%%%%%%%%%%%%%%%%%%%%%%%%%%%%%%%

\section{Experimental considerations}
The hexatic phase and other aspects of melting for aligned 2d dipolar systems have been observed with great precision  in the classical regime in experiments 
using colloidal particles with a  magnetic moment, confined to  an air-water interface.~\cite{Gasser}
On the other hand, the quantum analogue of  the hexatic phase is yet to be explored. 
The cold  assemblies  of dipolar molecules seem well suited to study both the classical and the quantum regimes of 2d melting. The typical dipole moment 
of these molecules is of the order of one Debye.
Taking as an example the recently  trapped  $^{23}$Na$^{40}$K molecule which has a  
permanent dipole moment of $d=2.7$ Debye,~\cite{Wu,DipoleMoment}
one gets  $r_D\simeq 24$ for an average inter particle spacing of $300$nm, which is well inside the quantum regime
as can be seen from Fig.~\ref{DipolePhase}. 
The critical temperature for the hexatic phase in the classical regime is $T\simeq 0.2\mu$K for this set of parameters. 
Even though the critical temperature will be lower in the quantum regime 
we estimate that  the quantum hexatic phase should be within experimental reach once the cooling techniques for the 
dipolar gases have been optimised. One can furthermore reach much higher critical temperatures using molecules with larger dipole moments 
such as SrO with $d=8.9$ Debye.  The presence of the hexatic phase can be detected by measuring the static structure factor via Bragg spectroscopy
which is a well proven experimental probe for quantum gases.~\cite{StamperKurn,Steinhauer,Veeravalli}
In the hexatic phase, the structure factor will exhibit a six fold symmetry with no sharp peaks similar to what it is shown in Fig.\ \ref{ExperimentalFig}. 
Recent impressive experiments have reported single atom resolution in optical lattices,~\cite{Sherson,Bakr} and if
 one is able to achieve the same  resolution with dipolar systems, the hexatic phase can be seen
 directly by the characteristic presence of  lattice defects consisting 
 of tightly bound disclination pairs, i.e., particles with 5 and 7 neighbors respectively.~\cite{NelsonBook}

\section{Conclusions}
In this paper, we analysed the stability of the crystal and hexatic phases of 2d systems consisting of 
either dipoles or charges.  
The classical elastic coefficients were calculated from the phonon spectra of the triangular crystal, and we then demonstrated 
how quantum effects decrease these coefficients thereby softening the crystal. Using Lindemann criteria suitably adapted to 
deal with the large fluctuations in 2d systems, we calculated approximate phase diagrams for the existence of the hexatic 
and crystal phases, predicting  that the hexatic phase is stable to very low temperatures. 
The relevant Lindemann numbers were extracted from experiments in the classical regime, and 
from Monte-Carlo calculations for $T=0$. The Lindemann numbers depend strongly on temperature, but they turn out to be 
 essentially the same for the 
charged and the dipolar system for the same value of $n_0\Lambda_T^2$. This suggests that the two-step melting of the crystal phase with an intermediate 
hexatic phase is a geometric phenomenon, insensitive to the detailed form of the particle interaction. 
Finally, we discussed the exciting prospect 
of finally being able to probe the existence hexatic phase in the quantum regime using ultra-cold dipolar gases.

\section{Acknowledgements}
It is a pleasure to acknowledge the Kavli Institute for Theoretical Physics, UCSB, where this work began. DRN would like to acknowledge
the hospitality of the Niels Bohr Institute, Copenhagen, and the support of the National Science Foundation, through grant DMR 1306367 and through 
the Harvard Material Research Science and Engineering Center via grant DMR-0820484. GMB would like to acknowledge the support of 
the Carlsberg Foundation via grant 2011\underline{\hspace{0.1cm}}01\underline{\hspace{0.1cm}}0264 and 
the Villum Foundation via grant VKR023163.
\appendix

\section{Ewald summation}\label{Ewald}
The phonon modes are as usual found by solving the matrix equation~\cite{AshcroftMermin} 
\begin{equation}
m\omega^2{\mathbf \epsilon}({\mathbf k})={\mathbf D}({\mathbf k}){\mathbf \epsilon}({\mathbf k})
\end{equation}
where ${\mathbf D}({\mathbf k})=\sum_{\mathbf R}{\mathbf D}({\mathbf R})\exp(-{\mathbf k}\cdot{\mathbf R})$ is the dynamical matrix with ${\mathbf R}$ 
the lattice vectors. We have 
\begin{equation}
{\mathbf D}_{ij}({\mathbf R})=E_0\times\begin{cases}
\sum_{{\mathbf R}\neq 0}\left[(n+2)n\frac{R_iR_j}{R^{n+4}}-n\frac{\delta_{ij}}{R^{n+2}}\right]& \mathbf{R}=0\\
n\frac{\delta_{ij}}{R^{n+2}}-(n+2)n\frac{R_iR_j}{R^{n+4}}& \mathbf{R}\neq0
\end{cases}
\end{equation}
for repulsive power law potentials $V(r)=E_0/r^n$
with $n=1$ and $E_0=Q^2$ for the charged particles, and $n=3$ and $E_0=D^2$ for the dipoles. We can write ${\mathbf D}({\mathbf k})$ as 
\begin{equation}
{\mathbf D}_{ij}({\mathbf k})=E_0\lim_{u\rightarrow 0}\frac{\partial^2}{\partial u_i\partial u_j}\sum_{{\mathbf R}\neq 0}\frac1{|{{\mathbf R}+{\mathbf u}}|^n}(1-e^{-i{\mathbf k}\cdot{\mathbf R}}).
\label{EwaldSum}
\end{equation}
Using $r^{-n}=(n+1)\pi^{-1/2}\int_0^\infty dy y^{n-1}e^{-r^2y^2}$ with $n=1,3$ and splitting the integral 
into a short range and a long range part, i.e.\ $\int_0^\infty dy\ldots=\int_0^{y_0} dy\ldots+\int_{y_0}^\infty dy\ldots$, the sum in Eq.\ (\ref{EwaldSum}) is split into a 
short range and a long range part,  ${\mathbf D}({\mathbf k})={\mathbf D}^<({\mathbf k})+{\mathbf D}^>({\mathbf k})$. Upon defining the function 
\begin{equation}
\varphi_n(x)=\frac 2{\sqrt\pi}\int_1^\infty dtt^ne^{-tx^2}
\end{equation}
we get 
\begin{gather}
{\mathbf D}_{ij}^>({\mathbf k})=E_0\frac{n+1}{2}y_0^{n+2}\sum_{{\mathbf R}\neq 0}(1-e^{-i{\mathbf k}\cdot{\mathbf R}})\nonumber\\
\times\left[2y_0^2R_iR_j\varphi_{n/2+1}(y_0R)-
\delta_{ij}\varphi_{n/2}(y_0R)\right].
\label{EwaldSumMore}
\end{gather}
The short range part of the sum is evaluated by Fourier transforming. Writing 
$\sum_{{\mathbf R}\neq 0}(1-e^{-i{\mathbf k}\cdot{\mathbf R}})\exp(-|{\mathbf R}+{\mathbf u}|^2y^2)=F_1({\mathbf u},y)-\exp({\mathbf k}\cdot{\mathbf u})F_2({\mathbf u},y)$
and Fourier transforming the functions $F_i({\mathbf u},y)$ which have the same periodicity as the lattice yields after some algebra
\begin{gather}
{\mathbf D}_{ij}^<({\mathbf k})=E_0\frac{n+1}{4\pi v}y_0^{n-2}\sum_{{\mathbf K}}
\left[({\mathbf K}+{\mathbf k})_i({\mathbf K}+{\mathbf k})_j\right.\nonumber\\
\left.
\times\varphi_{-n/2}(|{\mathbf K}+{\mathbf k}|/2y_0)-
{\mathbf K}_i{\mathbf K}_j\varphi_{-n/2}(|{\mathbf K}|/2y_0)
\right]
\label{EwaldSumLess}
\end{gather}
where $v=\sqrt 3 a^2/2$ is the area of the primitive cell of the lattice, and ${\mathbf K}$ are reciprocal lattice vectors. 
With Eqs.\ (\ref{EwaldSumMore})-(\ref{EwaldSumLess}) we have split 
the expression for ${\mathbf D}({\mathbf k})$
into two fast converging sums. These expressions agree with what is found in Refs.\ \onlinecite{Bonsall,Mora}. 
We pick  $y_0=1/a$ for the numerical calculations.  

\section{Canonical quantisation of the phonons}\label{Canonical}
For clarity, we briefly discuss  how quantum effects on the Lam\'e coefficients are included via 
canonical quantisation. In this approach, we introduce the  bosonic
annihilation operators $\hat b_{{\mathbf k}\sigma}$  for the phonons via 
\begin{equation}
\hat u_\sigma(\mathbf k)=\frac L{\sqrt{2\rho\omega_\sigma({\mathbf k})}}(\hat b_{{\mathbf k}\sigma}+\hat b^\dagger_{-{\mathbf k}\sigma})
\end{equation}
where $\sigma=l,t$.
Using $\langle\hat b^\dagger_{{\mathbf k}\sigma}\hat b_{{\mathbf k}\sigma}\rangle=[e^{\beta\hbar\omega_\sigma({\mathbf k})}-1]^{-1}$, we obtain
\begin{gather}
\langle \hat u_\sigma({\mathbf k})\hat u_\sigma(-{\mathbf k})\rangle=\frac{L^2}{\rho\omega_\sigma({\mathbf k})}(\frac{1}{e^{\beta\omega_\sigma({\mathbf k})}-1}+\frac 1 2)
\nonumber\\
=\frac{L^2}{2\rho\omega_\sigma({\mathbf k})}\coth[\beta\omega_\sigma({\mathbf k})/2)]
\end{gather}
which is identical to Eq.\ (\ref{Quantumuu}).


\begin{thebibliography}{31}
\bibitem{Jose} See articles in Jorge V.\ Jose (ed.), \textit{40 Years of Berezinskii-Kosterlitz-Thouless Theory} (World Scientific, Singapore 2013). 
\bibitem{Feynman} Feynman's theory is described in R.\ L.\ Elgin and D.\ L.\ Goodstein, in \textit{Monolayer and Submonolayer Films}, 
edited by J.\ G.\ Daunt and E.\ Lerner (Plenum, New York, 1973), and R.\ L.\ Elgin and D.\ L.\ Goodstein, Phys.\ Rev.\ A \textbf{9}, 2657 (1974).
\bibitem{NelsonBook}D.\ R.\ Nelson, \textit{Defects and Geometry in Condensed Matter Physics} (Cambridge University Press, 2002) and references therein. 
\bibitem{Gasser}U.\ Gasser, C.\ Eisenmann, G.\ Maret, and P.\ Keim, Chem.\ Phys.\ Chem.\ \textbf{11}, 963 (2010). 
\bibitem{Strandburg}See also C.\ M.\ Murray in  \textit{Bond-Orientational Order in Condensed Matter Systems (Partially Ordered Systems)} (Springer, Berlin 1992). 
\bibitem{Wu}C.-H.\ Wu, J.\ W.\ Park, P.\ Ahmadi, S.\ Will, and M.\ W.\ Zwierlein,  Phys.\ Rev.\ Lett. \textbf{109}, 085301 (2012).
\bibitem{Ni} K.-K.\ Ni, S.\ Ospelkaus, M.\ H.\ G.\ de Miranda, A. PeÕer, B.\ Neyenhuis, J.\ J.\ Zirbel, S.\ Kotochigova, 
P.\ S.\ Julienne, D.\ S.\ Jin, and J.\ Ye, Science \textbf{322}, 231 (2008).
\bibitem{Chotia} A.\ Chotia, B.\ Neyenhuis, S.\ A.\ Moses, B.\ Yan, J.\ P.\ Covey, M.\ Foss-Feig,
 A.\ M.\ Rey, D.\ S.\ Jin, and J.\ Ye, Phys.\ Rev.\ Lett.\ \textbf{108}, 080405 (2012).
\bibitem{Heo} M.-S.\ Heo, T.\ T.\ Wang, C.\ A.\ Christensen, T.\ M.\ Rvachov, D.\ A.\ Cotta, J.-H.\ Choi, Y.-R.\ Lee, 
 and W.\ Ketterle, Phys.\ Rev.\ A \textbf{86}, 021602(R) (2012).
 \bibitem{Pasquiou}B.\ Pasquiou, A.\ Bayerle, S.\ Tzanova, S.\ Stellmer, J.\ Szczepkowski, M.\ Parigger, R.\ Grimm, and F.\ Schreck,
 Phys.\ Rev.\ A \textbf{88}, 023601 (2013).
 \bibitem{Astrakharchik}G.\ E.\ Astrakharchik, J.\ Boronot, I.\ L.\ Kurbakov, and Yu.\ E.\ Lozovik, Phys.\ Rev.\ Lett. \textbf{98}, 060405 (2005).
\bibitem{Buchler}H.\ P.\ B\"uchler,  E.\ Demler, M.\ Lukin, A.\ Micheli, N.\ ProkofÕev, G.\ Pupillo, and P.\ Zoller, Phys.\ Rev.\ Lett. \textbf{98}, 060404 (2005).
\bibitem{Grimes}C.\ C.\ Grimes and G.\ Adams, Phys.\ Rev.\ Lett.\ \textbf{42}, 795 (1979).
\bibitem{Morf}R.\ Morf, Phys.\ Rev.\ Lett.\ \textbf{43}, 931 (1979).
\bibitem{Muto}S.\ Muto and H.\ Aoki, Phys.\ Rev.\ B \textbf{59}, 14911 (1999); W.\ J.\ He, T.\ Cui, Y.\ M.\ Ma, Z.\ M.\ Liu, and G.\ T.\ Zou,
Phys.\ Rev.\ B \textbf{68}, 195104 (2003).
\bibitem{Runge}K.\ J.\ Runge and G.\ V.\ Chester, Phys.\ Rev.\ B \textbf{38}, 135 (1988). 
\bibitem{Babadi}M.\ Babadi, B.\ Skinner, M.\ M.\ Fogler, and E.\ Demler, Eur.\ Phys.\ Lett.\ \textbf{103}, 16002 (2013). 
\bibitem{Lindemann} F.\ Lindemann, Phys.\ Z.\ \textbf{11}, 609 (1910).
\bibitem{Blatter} For a review of recent applications to the melting of vortex matter in high $T_c$  superconductors, see G.\ Blatter and V.\ B.\ Geshkenbein, 
in \textit{Superconductivity, Vol 2: Novel Superconductors}, edited by K.\ H.\ Bennemann, J.\ B.\ Ketterson  pp. 495-637 (Springer, Berlin 2008). 
  See also, G.\ Blatter, M.\ V.\ Feigel'man, V.\ B.\ Geshkenbein, A.\ I.\ Larkin,and V.\ M.\ Vinokur, Rev.\ Mod.\ Phys.\ \textbf{66}, 1125 (1994).  
\bibitem{FisherFisher}D.\ S.\ Fisher, M.\ P.\ A.\ Fisher, and D.\ A.\ Huse, Phys.\ Rev.\ B \textbf{43}, 130 (1991).
\bibitem{LandauLifshitz}L.\ D.\ Landau and E.\ M.\ Lifshitz, \textit{Theory of Elasticity} (Elsevier, 2008). 
\bibitem{Mermin} N.\ D.\ Mermin, Phys.\ Rev.\ B \textbf{176}, 250 (1968). 
\bibitem{NelsonHalperin} D.\ R.\ Nelson and B.\ I.\ Halperin, Phys.\ Rev.\ B \textbf{19}, 2457 (1979).
\bibitem{Engel} The latter situation may arise in computer simulations of the melting of classical hard disks, which after 55 years(!) of effort, 
now finally point to a thin sliver of hexatic phase, separated from an isotropic fluid by a first order phase transition.  
 See M.\ Engel, J.\ A.\ Anderson, S.\ C.\ Glotzer, M.\ Isobe, E.\ P.\ Bernard, and W.\ Krauth, Phys.\ Rev.\ E \textbf{87}, 042134 (2013).
\bibitem{AshcroftMermin}N.\ W.\ Aschcroft  and N.\ D.\ Mermin, \textit{Solid State Physics} (Saunders Company, 1976). 
\bibitem{Clark}B.\ K.\ Clark, M.\ Casula, and D.\ M.\ Ceperley, Phys.\ Rev.\ Lett. \textbf{103}, 055701 (2009).
\bibitem{FeynmanHibbs}R.\ P.\ Feynman and A.\ R.\ Hibbs, \textit{Quantum Mechanics and Path Integrals} ( McGraw-Hill , 1965). 
 \bibitem{Reppy}J.\ D.\ Reppy, Phys.\ Rev.\ Lett.\ \textbf{104}, 255301 (2010).
\bibitem{Balibar}S.\ Balibar, Nature \textbf{464}, 176 (2010).  
\bibitem{Oganesyan} V.\ Oganesyan, S.\ A.\ Kivelson and Eduardo Fradkin, Phys.\ Rev.\ B \textbf{64} 195109 (2001); B.\ M.\ Fregoso and E.\ Fradkin, 
   Phys.\ Rev.\ Lett.\ \textbf{103}, 205301 (2009).
\bibitem{Mullen} K.\ Mullen, H.\ T.\ C.\ Stoof, M.\ Wallin, and S.\ M.\ Girvin Phys.\ Rev.\ Lett.\ \textbf{72}, 4013 (1994).
\bibitem{NelsonSeung} D.\ R.\ Nelson and H.\ S.\ Seung, Phys.\ Rev.\ B \textbf{39}, 9153 (1989).
\bibitem{Marchetti}  M.\ C.\ Marchetti and D.\ R.\ Nelson, Phys.\ Rev.\ B \textbf{41}, 1910 (1990).
\bibitem{Strey}H.\ H.\ Strey, J.\ Wang, R.\ Podgornik, A.\ Rupprecht, L.\ Yu, V.\ A.\ Parsegian, and E.\ B.\ Sirota, Phys.\ Rev.\ Lett.\ \textbf{84}, 3105 (2000).
\bibitem{Lu}X.\ Lu, C.-Q.\ Wu, A.\ Micheli, and G.\ Pupillo, Phys.\ Rev.\ B \textbf{78}, 024108 (2008).
\bibitem{Grunberg}  H.\ H.\ von Gr\"unberg, P.\ Keim, and G.\ Maret, Phys.\ Rev.\ Lett. \textbf{93}, 255703 (2004).
\bibitem{Fisher}D.\ S.\ Fisher, B.\ I.\ Halperin, and R.\ Morf, Phys.\ Rev.\ B \textbf{20}, 4692 (1979).
\bibitem{ChaikinBook}P.\ M.\ Chaikin and T.\ C.\ Lubensky, \textit{Principles of Condensed Matter Physics} (Cambridge University Press, 1995). 
\bibitem{MerminWagner}N.\ D.\ Mermin and H.\ Wagner, Phys.\ Rev.\ Lett. \textbf{17}, 1133 (1966).
\bibitem{HalperinNelson} B.\ I.\ Halperin and D.\ R.\ Nelson, Phys.\ Rev.\ Lett.\ \textbf{41}, 121 (1978).
\bibitem{Kalia}R.\ K.\ Kalia and P.\ Vashishta, J.\ Phys.\ C \textbf{14}, L643 (1981).
\bibitem{Lozovik2}Yu.\ E.\ Lozovik and  V.\ M.\ Fartzdinov, Solid State Comm.\ \textbf{54}, 725 (1985).
\bibitem{DipoleMoment} The effective dipole moment of the molecules will be smaller 
 in an experiment due to molecule rotation. 
 \bibitem{StamperKurn}D.\ M.\ Stamper-Kurn, A.\ P.\ Chikkatur, A.\ G\"orlitz, S.\ Inouye, S.\ Gupta, D.\ E.\ Pritchard, and W.\ Ketterle,
Phys.\ Rev.\ Lett.\ \textbf{83}, 2876 (1999).
\bibitem{Steinhauer}J.\ Steinhauer, R.\ Ozeri, N.\ Katz, and N.\ Davidson, Phys.\ Rev.\ Lett.\ \textbf{88}, 120407 (2002).
\bibitem{Veeravalli}G.\ Veeravalli, E.\ Kuhnle, P.\ Dyke, and C.\ J.\ Vale, Phys. Rev. Lett. 101, 250403 (2008).
\bibitem{Sherson}J.\ F.\ Sherson, C.\ Weitenberg, M.\ Endres, M.\ Cheneau, I.\ Bloch, and S\ Kuhr, Nature \textbf{467}, 68 (2010).
\bibitem{Bakr}W.\ S.\ Bakr, J.\ I.\ Gillen, A.\ Peng, S.\ F\"olling, and  M.\ Greiner, Nature \textbf{462}, 74 (2009).
\bibitem{Bonsall}L.\ Bonsall and A.\ A.\ Maradudin, Phys.\ Rev.\ B \textbf{15}, 1959 (1977).
\bibitem{Mora}C.\ Mora, O.\ Parcollet, and X.\ Waintal, Phys.\ Rev.\ B \textbf{76}, 064511 (2007).


\end{thebibliography}
\end{document}